\documentclass[twocolumn,english,aps]{revtex4-1}
\usepackage[T1]{fontenc}
\usepackage[latin9]{inputenc}
\usepackage{float}
\usepackage{amsmath}
\usepackage{amssymb}
\usepackage{graphicx}
\usepackage{esint}

\makeatletter

\usepackage{babel}

\usepackage{babel}

\usepackage{babel}

\usepackage{hyperref}

\makeatother

\usepackage{babel}
\begin{document}
	
	\title{Kitaev quasiparticles in a proximate spin liquid: A many-body localization
		perspective}
	
	\author{Aman Kumar}
	
	\affiliation{Department of Theoretical Physics, Tata Institute of Fundamental
		Research, Homi Bhabha Road, Navy Nagar, Mumbai 400005, India}
	
	\author{Vikram Tripathi}
	
	\affiliation{Department of Theoretical Physics, Tata Institute of Fundamental
		Research, Homi Bhabha Road, Navy Nagar, Mumbai 400005, India}
	
	\date{\today}
	\begin{abstract}
		We study the stability of Kitaev quasiparticles in the presence of
		a perturbing Heisenberg interaction as a Fock space localization phenomenon.
		We identify parameter regimes where Kitaev states are localized, fractal
		or delocalized in the Fock space of exact eigenstates, with the first two
		implying quasiparticle stability. Finite temperature calculations
		show that a vison gap, and a nonzero plaquette Wilson loop at low
		temperatures, both characteristic of the deconfined Kitaev spin liquid
		phase, persist far into the neighboring phase that has a concomitant stripy 
		spin-density wave (SDW) order. The key experimental implication 
		for Kitaev materials is that below a characteristic energy scale, unrelated to the SDW ordering, 
		Kitaev quasiparticles are stable.
	\end{abstract}
	\maketitle
	The honeycomb Kitaev model describes an integrable $Z_{2}$ quantum
	spin liquid, with quasiparticles consisting of gapped $Z_{2}$
	fluxes (visons) and deconfined Majorana fermions (spinons) \cite{Kitaev:2006lla}.
	Considerable theoretical \cite{gohlke2017dynamics,winter2017breakdown,gordon2019theory,Khaliullin,Kitaev-Heisenbeeg-J2_J3_PhysRev,katukuri2014kitaev,Sizyuk_Phys_iridates,annurev-conmatphys-031115-011319,Youhei_iridates_PhysRevLett,bhattacharjee2012spin,Hu_Iridates_PhysRevLett.115.167204,quantum_hall_shitade_PhysRevLett.102.256403,Khaliullin_iridates_PhysRevB.92.024413,Takayama_iridates_PhysRevLett.114.077202,kim2014antiferromagnetic,Zigzag_khaliullin_PhysRevLett.110.097204,Liu2018Dirac,Raman_Scattering_moessner_2014,hickey2019emergence}
	and experimental \cite{banerjee2016,do2017majorana,WangTHz2017,yu2018,J_K_model_exp_PhysRevLett.108.127203,zigzag_phase_iridates_PhysRevB.85.180403,Spin_waves_Choi_PhysRevLett.108.127204,AFM_iridates_PhysRevB.82.064412,chun2015direct,Comin_2012_PhysRevLett.109.266406,gretarsson2013crystal,gretarsson2013magnetic,liu2011long,mehlawat2017heat,banerjee2018excitations,magnons_THZ,matsuda2018majorana}
	debate surrounds the question of realizing Kitaev physics in the
	presence of competing spin-interactions, since the ground state in
	Kitaev materials \cite{inelastic_neutron,Sandilands_kitaev_candidates,Sandilands_luke_kitaev_candidates,Plumb_iridates,Jackeli_Mott},
	is magnetically ordered. Data from inelastic neutron scattering \cite{inelastic_neutron,winter2017breakdown},
	THz spectroscopy \cite{WangTHz2017,magnons_THZ}, thermal conductivity \cite{yu2018,Leahy2017_thermalconductivity_torque},
	thermal Hall response \cite{matsuda2018majorana}, and high field torque
	magnetometry \cite{Leahy2017_thermalconductivity_torque,torque_responce}
	have been interpreted in both ways \textendash{} as evidence of Kitaev
	physics \cite{matsuda2018majorana,gordon2019theory,torque_responce,banerjee2016,baek2017evidence}
	as well as magnon interaction \cite{winter2017breakdown,kim2014antiferromagnetic,magnons_THZ,liu2011long,gretarsson2013magnetic}. 
	This motivates a basic question: can many-body
	excitations of a Kitaev model subjected to a Heisenberg perturbation
	resemble Kitaev quasiparticles even as the ground state may have magnetic order? 
	
	Here we study Kitaev quasiparticle stability for a simple ($J-K)$ model, consisting of ferromagnetic
	(FM) Kitaev ($K$) and antiferromagnetic (AFM) Heisenberg ($J$) interactions,
	as a Fock space localization problem. We use state-of-the-art exact diagonalization methods, namely FEAST
	\cite{FEAST} and  Krylov-Schur \cite{krylov_schur}, for systems of up to $N=30$ spins. 
	The problem is also interesting for understanding Fock space localization transitions in disorder-free interacting 
	systems \cite{Moessner_disorder_free_localization,Moessner_Absence_ergodicity,Brenes_gauge_invariance_mbl,diamantini2018confinement},
	and shedding light on the question of existence of one or two localization transitions \cite{altshuler1997, mirlin1997}.
	
	The ground state of our model is known \cite{Khaliullin,Zigzag_khaliullin_PhysRevLett.110.097204}
	to be a Kitaev spin liquid (KSL) for $0\leq J/K\lesssim0.12,$ stripy AFM for $0.12\lesssim J/K\lesssim0.75,$
	and N\'{e}el AFM for larger $J/K.$ The stripy order peaks at $J/K=0.5.$ Here a sublattice transformation \cite{Khaliullin} maps the model exactly to an 
	isotropic Heisenberg ferromagnet establishing stripy AFM as the exact, direct product ground state at this point. The
	regime $0.12\lesssim J/K\lesssim0.5$ is what we will call a proximate spin liquid
	phase (PSL) where the ground state shows magnetic order but Kitaev 
	correlations are significant.
	
	We show that low-lying Kitaev quasiparticles are stable over a significant
	interaction parameter range in the PSL phase. In the PSL phase, the low-lying eigenstates, which have attributes of both magnons as well as Kitaev, resemble Kitaev quasiparticles better than magnons near the KSL/PSL boundary, and magnons near the peak of the stripy order at $J/K=0.5.$ We also obtain a Kitaev stability curve, unrelated to the SDW ordering scale $T_N,$ below which Kitaev quasiparticles are stable. This includes regions both above and below $T_N,$ in contrast with recent work \cite{Rousochatzakis} that restricted the region of Kitaev stability to above $T_N$ in the vicinity of the KSL/PSL boundary. 
	Our finite temperature
	calculations show KSL signatures such as a vison gap, and at low temperatures,
	a nonzero value of the Kitaev fluxes, persisting in the PSL all the way to $J/K=0.5.$ These findings open the possibility of exploring Kitaev physics in a large window of energies reckoned from the ground state of Kitaev materials. We discuss our findings in the context of experiments, and propose new tests.

	The problem of quasiparticle stability in interacting systems is deeply connected to the many-body localization (MBL) phenomenon \cite{altshuler1997}.
	To see this, we represent individual Kitaev eigenstates as linear
	superpositions of the exact eigenstates of the $(J-K)$ model. 
	The scaling of the support size $\xi$ of the Kitaev states with the
	dimensionality $D=2^{N}$ of the Fock space as $\xi\sim D$ implies a fully many-body delocalized state or a
	decaying quasiparticle, while $\xi\sim D^{0}$ corresponds to a localized
	state or non-decaying quasiparticle. A third possibility,
	$\xi\sim2^{cN}=D^{c},$ with $c<1,$ represents a fractal delocalized
	state and still corresponds to a long-lived quasiparticle excitation
	since $\xi/D\rightarrow0$ as $D\rightarrow\infty.$
	
	We begin our analysis with the following nearest-neighbour $J-K$
	model: 
	\begin{align}
	H & =-K\sum_{\langle ij\rangle,\gamma}\sigma_{i}^{\gamma}\sigma_{j}^{\gamma}+J\sum_{\langle ij\rangle}\boldsymbol{\sigma}_{i}\cdot\boldsymbol{\sigma}_{j},\label{eq:model}
	\end{align}
	where $K,J>0,$ $\gamma=x,y,z$ labels an axis in spin space and a
	bond direction of the honeycomb lattice and $\sigma_{i}^{\gamma}$
	represent Pauli spin matrices at the site labeled $i.$ We consider only
	clusters with (even) number $N$ of spins ranging from 12 to 30, where toroidal periodic boundary conditions
	can be applied.
	
	We use the FEAST eigensolver algorithm \cite{FEAST} that allows computation of
	large numbers of eigenvectors, including degeneracies, in arbitrarily
	specified energy ranges,
	and is parallelizable at multiple stages. 
	The FEAST algorithm uses the contour integration based projector, 
	\begin{align}
	\frac{1}{2\pi i}\oint_{C} & \frac{dE}{EI-H}|v\rangle=\sum_{n\in C}\langle n|v\rangle|n\rangle,\label{eq:feast}
	\end{align}
	where $|v\rangle$ is in general some random vector defined on the
	entire Fock space of dimension $D=2^{N},$ and $\{|n\rangle\}$ are
	the eigenvectors corresponding to, say, $m$ eignenvalues \cite{di2016efficient} lying within
	the user-defined contour $C$. By choosing a number $p\geq m,$ of
	these random vectors (in general linearly-independent), we end up
	with a set of $m$ linearly-independent vectors spanning the eigenspace
	enclosed in $C.$ For evaluation of the lowest eigenstates, we interchangeably
	used the computationally cheaper and faster Krylov-Schur algorithm \cite{krylov_schur}, 
	which yields around 1000 low-lying eigenstates
	for each Bloch momentum \cite{sandvik2010computational} for $N=30$ on our machine and suffices to demonstrate
	Fock space localization of low-lying Kitaev states.
	
	To study the resemblance of a given Kitaev state $|\alpha_{k}\rangle$
	with the exact eigenstates $|\psi_{i}\rangle$ of the $J-K$ model,
	we expand it as a linear superposition, 
	\begin{align}
	|\alpha_{k}\rangle & =\sum_{i=1}^{D}a_{ki}|\psi_{i}\rangle,\label{eq:linsup}
	\end{align}
	and obtain the inverse participation ratio (IPR), 
	\begin{align}
	P_{k} & =\sum_{i=1}^{D}|a_{ki}|^{4}.\label{eq:ipr-def}
	\end{align}
	The support size for the state $|\alpha_{k}\rangle$ is then $\xi_{k}=1/P_{k}.$
	In practice, only the states contributing significantly to the IPR
	need be computed - these states have relatively large overlaps
	$a_{ki}.$ 
	For comparison with stripy states, we also perform
	the same analysis for eigenstates corresponding to $J/K=0.5.$
	The nature of localization of a state in Fock space is determined
	from a finite size scaling analysis of the support sizes.
	
	\begin{figure}
		\includegraphics[width=1\linewidth]{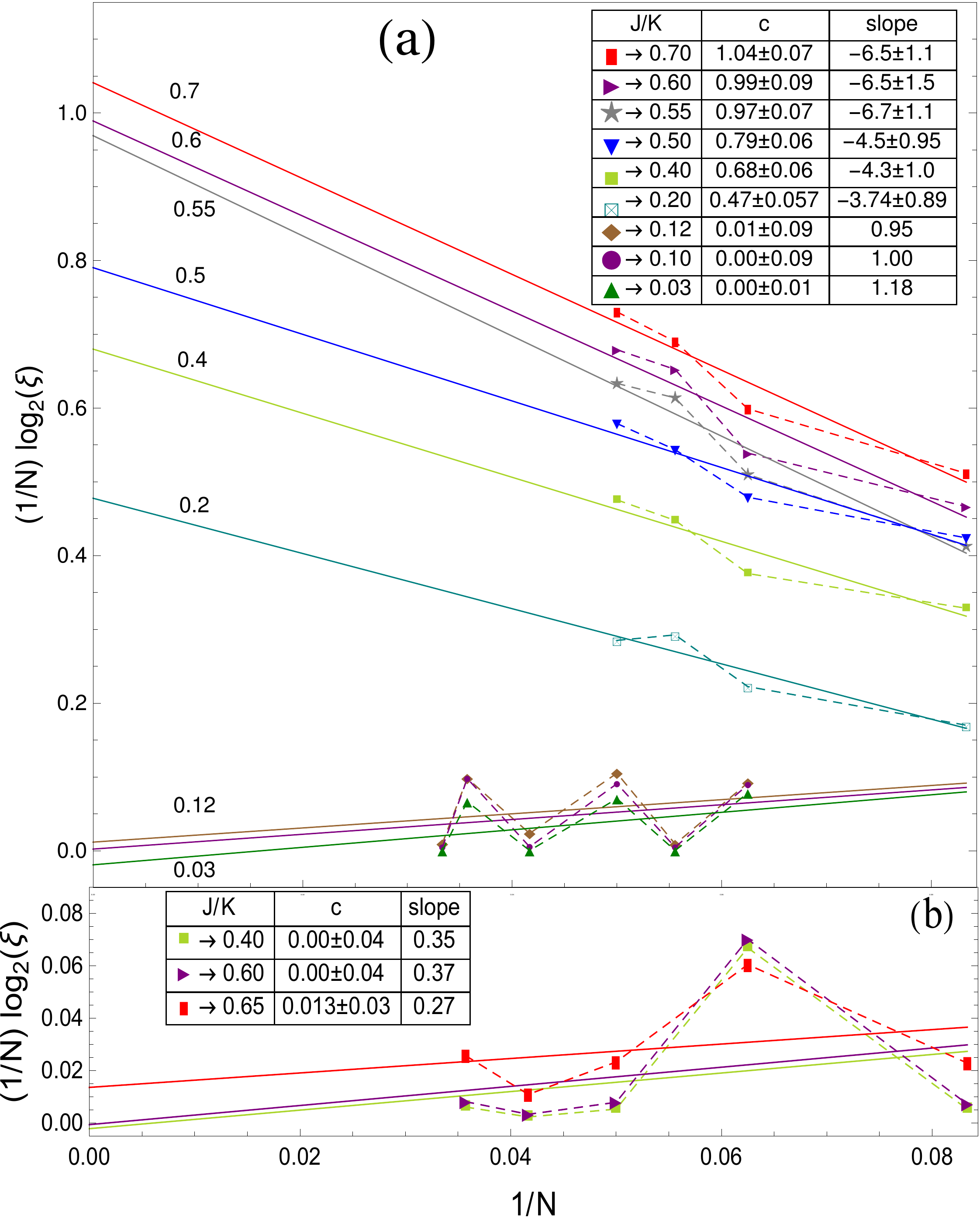}
		\caption{\label{fig:N-scaling}Plots showing finite size scaling of the support
			size, $\xi,$ of (a) a low energy Kitaev state (lowest two vortex state) and (b) the stripy ground state for $J/K=0.5,$ in the Fock space of eigenstates of the $J-K$ model for up to $N=30$ spins. The fits (solid lines)
			are to a law of the form $\xi=f2^{cN}.$ Lines with negative (positive)
			slopes correspond to many-body delocalized (localized) phases. The numbers on the solid
			lines indicate the corresponding values of $J/K.$ In (a), a 
			localized ($c=0$) to fractal ($0<c<1$) transition occurs at the Kitaev-stripy
			AFM phase boundary, $J/K\approx0.12$ and a fractal to fully delocalized ($c=1$) beyond
			the peak of the stripy order, $J/K=0.5.$ Full Kitaev delocalization is further indicated in (b) where the SDW ground state shows localization for $0.4 < J/K < 0.65.$}
	\end{figure}
	
	\begin{figure*}
		\includegraphics[width=0.7\columnwidth]{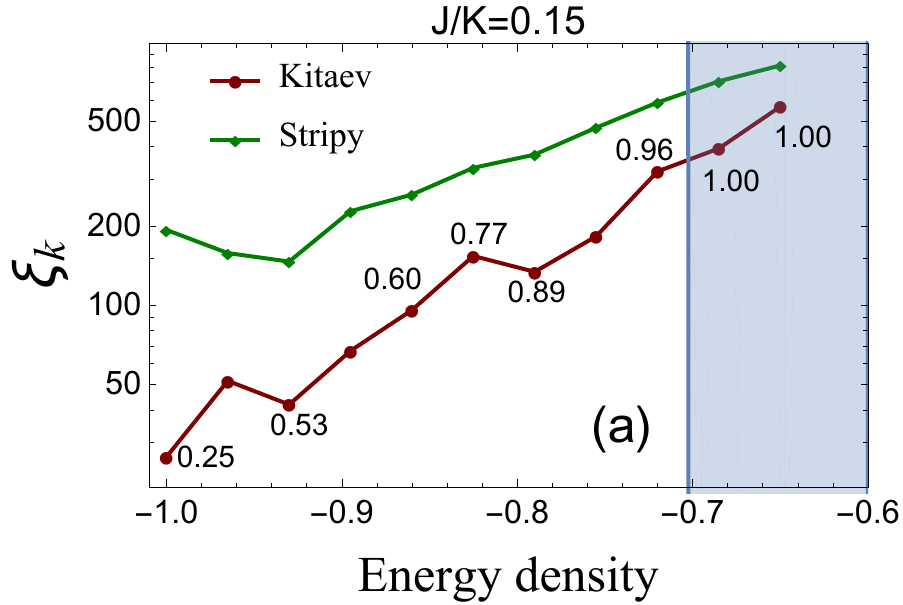}
		\includegraphics[width=0.64\columnwidth]{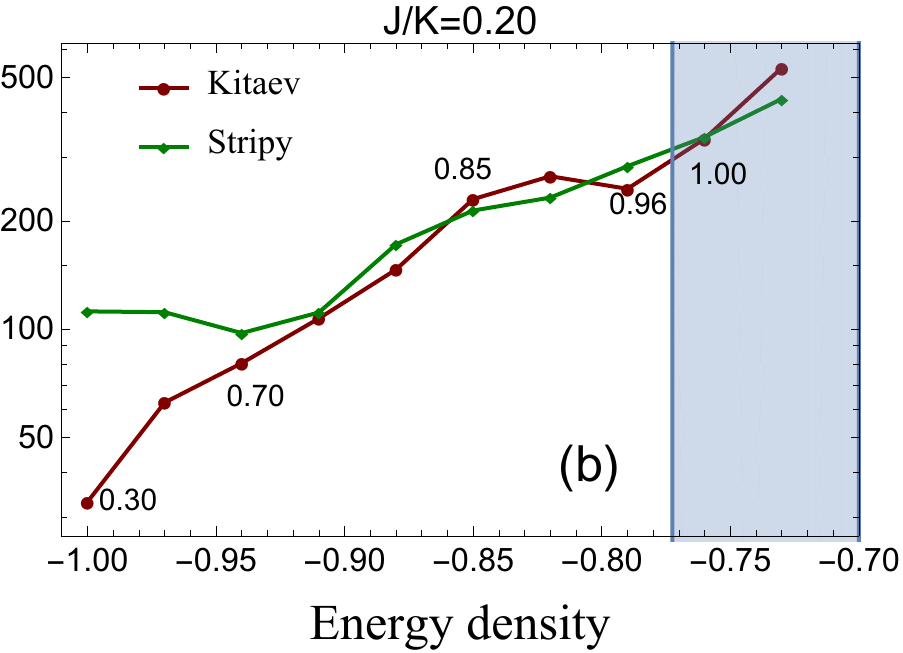}
		\includegraphics[width=0.63\columnwidth]{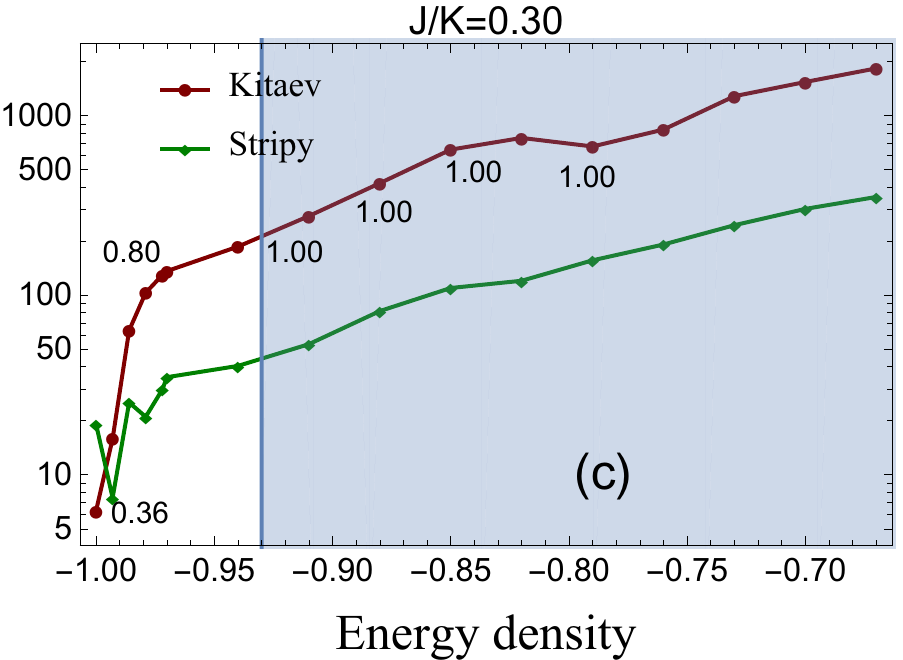}
		\caption{\label{fig:resemblance}Calculated support sizes $\xi$ for the Kitaev
			and stripy SDW states (corresponding to $J/K=0$ and $J/K=0.5$ respectively)
			for three different points in the proximate spin liquid (PSL) phase as a function of energy density. 
			The numbers on the curves show the values of the exponent $c$ for
			the Kitaev states obtained from finite-size scaling. In the
			white regions, Kitaev states show fractal finite size scaling,
			$\xi\sim2^{cN},\,c<1,$ (stable Kitaev quasiparticles) while the (blue)
			shaded regions correspond to fully delocalized (unstable) Kitaev states.
			The exact eigenstates
			in the PSL are more Kitaev-like at low energy densities and smaller
			values of $J/K.$}
	\end{figure*}
	
	Figure \ref{fig:N-scaling} shows plots of $(1/N)\log_{2}\xi$ vs $1/N$
	for (a) the lowest two-flux Kitaev state (a low-lying state) and (b) the direct product ground state at 
	the peak of stripy order, $J/K=0.5,$ for $N$ up to 30, together with linear fits.
	In the fully delocalized or fractal phases, where $\xi\sim f2^{cN},$ the
	slope ($-\ln_{2}(1/f)$) is negative and $c>0,$ while in the MBL phase, the slope is positive. 
	The intercept yields
	the exponent $c.$ In the KSL phase, $J/K\lesssim0.12,$ the support $\xi$ is $O(1)$ even for $N=30$, 
	the slope positive, and $c\approx 0.$ 
	The large data scatter in the KSL reflects $O(1)$ changes from sample to sample, but the trend is clear. The KSL phase is known to be a spin-liquid with Kitaev quasiparticles.
	For $0.12\lesssim J/K\lesssim 0.5,$ a fractal delocalized phase,
	the exponent smoothly rises to $c\approx0.8.$ A small increase to $J/K=0.55$ results in a sharp relative increase to $c\approx1,$
	signifying full delocalization. These MBL transitions respectively occur in 
	the vicinity of well-known points - the Kitaev-stripy transition point ($J/K\approx0.12$) and at $J/K=0.5.$
	Delocalization of the Kitaev states is also indicated in (b) in the window $0.4<J/K<0.65,$ where the SDW is localized reflecting the fact that the stripy states have small overlaps with the deconfined Kitaev states. Independent check for destabilization of Kitaev states beyond $J/K=0.5$ comes from the flux expectation values (see below). We also
	studied the scaling behavior of the entanglement entropy (see SM), $S_{k}  =-\sum_{i=1}^{D}|a_{ki}|^{2}\log_{2}|a_{ki}|^{2},$
	and reached conclusions consistent with scaling of $\xi.$ The entropy follows a volume law scaling in the fully delocalized phase, which interestingly is not an ergogic phase since the expectation value of local quantities like the flux fluctuates sharply between states with nearby energies.
	
	Figure \ref{fig:resemblance} shows support sizes for Kitaev
	($J/K=0$) and stripy AFM ($J/K=0.5$) states as a function of the
	energy density for three representative values of $J/K$ in the PSL
	phase for $N=18.$ The energy density is the energy per site in dimensionless units,
	obtained by normalizing the energies by the ground state energy, an extensive quantity.
	The choice of energy density is based on our observation
	that states with comparable energies have comparable support sizes
	(see SM). For $J/K=0.15,$ near the KSL
	boundary (Fig. \ref{fig:resemblance}(a)), Kitaev states
	have smaller support sizes compared to stripy SDW. The numbers on the
	Kitaev curve show the exponents $c$ obtained from finite-size
	scaling. The white (blue) shaded regions represent the fractal (delocalized)
	phase of Kitaev states.  The stripy SDW states for $J/K=0.15$ delocalize at lower energy densities than the corresponding
	Kitaev states. For $J/K=0.2$ (Fig. \ref{fig:resemblance}(b)), Kitaev states are still more localized than stripy SDW at low energy densities. Thereafter the Kitaev and stripy SDW states  show similar scaling behavior.
	Further into the PSL phase (\ref{fig:resemblance}(c)), for $J/K=0.3,$ there is possibly a very small region of low energy
	density where Kitaev states are more stable than magnons. Here the magnons continue
	to show fractal scaling (stability) to much higher energy densities
	($\sim-0.7$). We conclude that in the PSL phase, the exact eigenstates
	may be approximated as Kitaev states for sufficiently low energy densities,
	and beyond a high enough energy density that depends on $J/K,$
	neither Kitaev and stripy SDW descriptions are appropriate.
	A phase diagram showing stability of Kitaev states in the $J/K$ vs. energy density plane is shown in the SM.
	
	\begin{figure}
		\includegraphics[width=1\columnwidth]{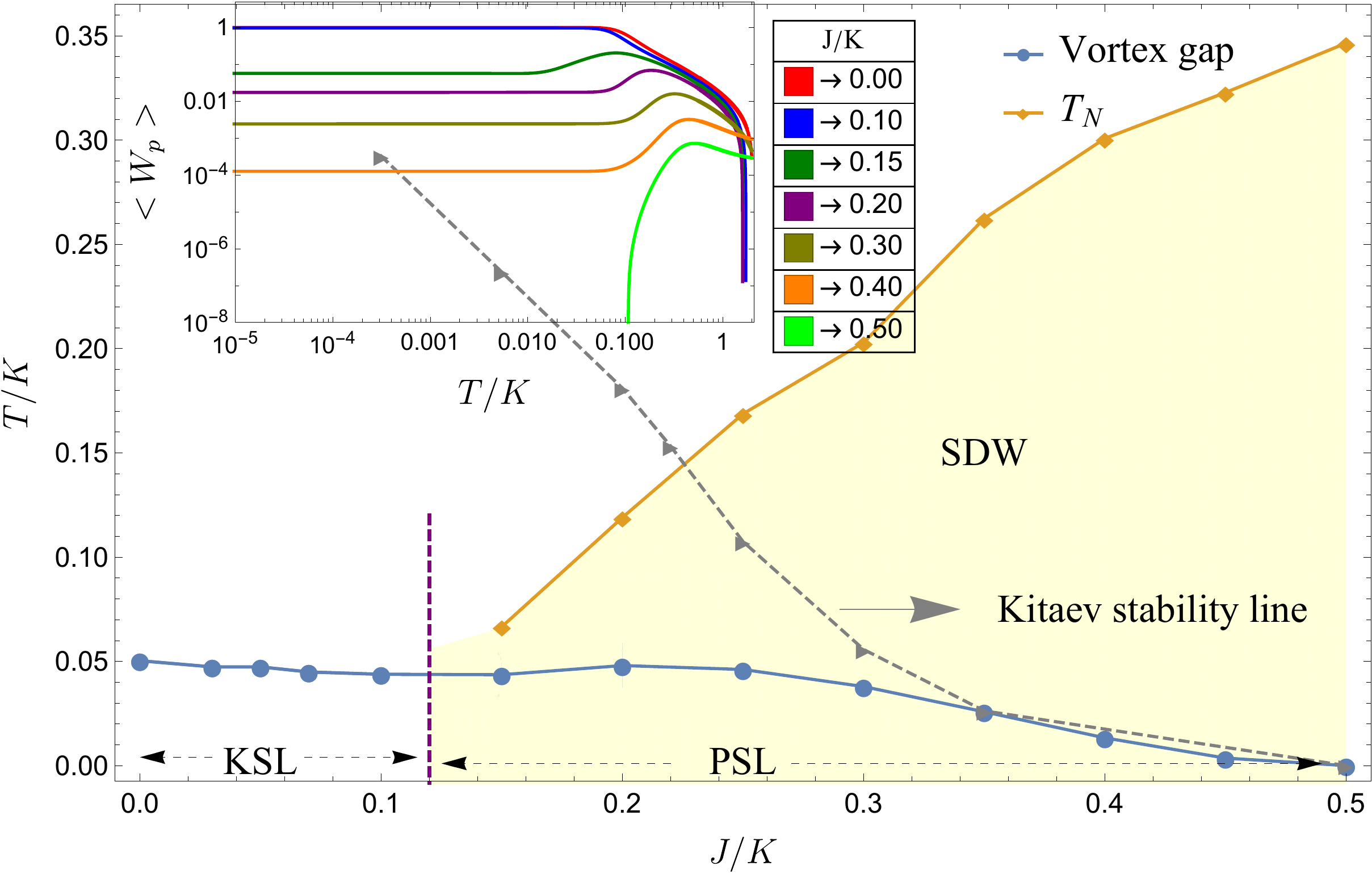}
		
		\caption{\label{fig:Tphase_diag}Finite temperature phase diagram for the $J-K$
			model obtained from specific heat calculations. The vison (vortex)
			gap, characteristic of the KSL phase, persists well into the PSL phase,
			vanishing only at $J/K=0.5$ where stripy SDW order peaks, and beyond.
			The Kitaev stability line, obtained from MBL calculations, marks a boundary
			below which Kitaev quasiparticles are stable. This includes regions above
			the magnetic ordering scale $T_N$ as well as parts of the SDW phase.
			The plaquette Wilson loop operators, or Kitaev fluxes have finite expectation values at temperatures
			lower than the vison gap. The inset shows
			the temperature dependence of the Kitaev fluxes $\langle W_{p}\rangle,$ for different $J/K.$}
	\end{figure}
	
	Figure \ref{fig:Tphase_diag} shows the temperature ($T$) vs. $J$
	phase diagram obtained from specific heat calculations for
	a 24-site cluster. The curves indicate onset of stripy SDW order
	at higher temperatures, the evolution of the vison gap, and the line of Kitaev quasiparticle stability.
	The vertical dashed line ($J/K\approx0.12$) marks the KSL/PSL phase transition, known to be of first order \cite{gohlke2017dynamics}. The vison gap, a distinguishing
	feature of the KSL phase, extends well into the PSL phase, vanishing
	only at $J/K=0.5$ where stripy SDW order peaks
	(and beyond). We also estimated the vison gap directly by introducing fluxes (see SM), which agree with those from specific heat calculations. The inset shows the Kitaev fluxes $W_{p}$
	as a function of temperature, for different $J/K.$ In the deconfined KSL phase at low temperatures, we find  
	$\langle W_{p}\rangle=1,$ implying that Heisenberg
	perturbations do not confine. 
	Entering the PSL phase, we find $\langle W_{p}\rangle$
	smoothly decreases with increasing $J/K$ but vanishes only at $J/K=0.5,$ where confinement occurs. The nonvanishing vison gap at $J/K=0.12$ is consistent with the first order nature of the SDW transition. Kitaev excitations are well-defined below the Kitaev stability curve (obtained from MBL calculations). The stability region includes parts of both within the SDW phase (shaded yellow) and regions above $T_N.$ While the latter is consistent with the understanding in Ref. \cite{Rousochatzakis}, the former, to our understanding, is new. Moreover we have shown that the low-lying excitations better resemble the Kitaev states than the highly excited ones such as those above $T_N.$
	
	\begin{figure}
		\includegraphics[width=1\columnwidth]{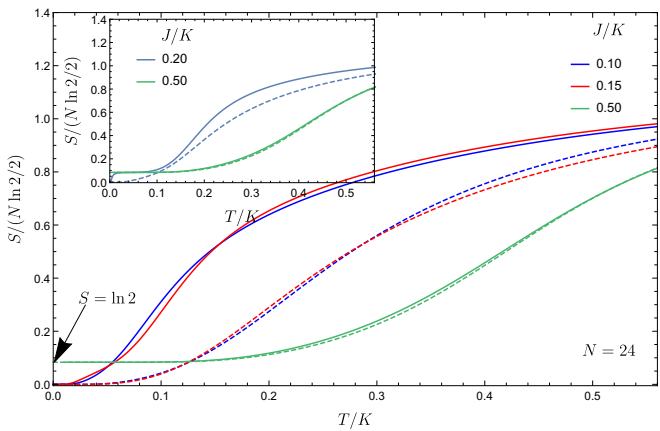}
		\caption{\label{fig:entropyT}Figure illustrating the sensitivity of the thermodynamic entropy to an external magnetic field in the $(1,1,1)$ direction in KSL and PSL regimes. Two values of $B$ are shown, $B/K=0.1\sqrt{3}$ (solid lines) and $0.2\sqrt{3}$ (dashed lines). Increasing magnetic field strongly suppresses the entropy in both the KSL ($J/K=0.1$) and PSL ($J/K=0.15,\,0.2$) regimes but leaves the direct product stripy state at $J/K=0.5$ unaffected. The $\ln(2)$ entropy at low temperatures for the direct product magnet reflects the two-fold degeneracy of the stripy state. 
		}
	\end{figure}
	
	In Fig. \ref{fig:entropyT} we show the temperature dependence of the thermodynamic entropy for different values of $J/K,$ at two different values of an external magnetic field ($B/K=0.1(1,1,1),\,0.2(1,1,1)$) -- a quantity that may be obtained from finite field specific heat measurements. In the vicinity of the KSL/PSL phase boundary for both $J/K=0.1$ in the KSL and $J/K=0.15,\,0.2$ in the PSL, the low-temperature entropy gets strongly suppressed by increasing the magnetic field, while the entropy of direct product magnetically ordered phase at $J/K=0.5$ is relatively unaffected.
	
	To conclude, Kitaev quasiparticles decay when
	perturbed by sufficiently strong Heisenberg interactions. By mapping the problem to one of many-body localization
	in the Fock space, we showed that over a range of $J/K$
	in the PSL phase, the low-lying excitations
	are more Kitaev-like than magnons of an SDW state. Crucially, we obtained an upper energy bound for the Kitaev quasiparticle stability. Finite temperature analysis of 
	the specific heat revealed a finite vison gap within the PSL phase. The thermodynamic entropy showed strong sensitivity to an external magnetic field on either side of the KSL/PSL boundary but not deep in the PSL.
	
	\textit{Experimental implications-} Our study shows that near the KSL/PSL phase boundary, low-lying excitations in the PSL phase are more likely to resemble Kitaev quasiparticles than magnons, a view supported by the low temperature observations of quantized thermal Hall conductivity \cite{matsuda2018majorana}  and 
	from high-field magnetometry \cite{torque_responce} of Kitaev materials. The upper (energy) bound on Kitaev stability that we obtained encompasses regions both above and below $T_N,$ with the former possible only near the KSL/PSL boundary. Thus the observation of incoherent features in neutron scattering at energies above $T_N$ \cite{inelastic_neutron,winter2017breakdown} in $\alpha$-RuCl$_3,$ may be a signature of Kitaev physics provided one is sufficiently close to the KSL phase boundary. Since the presence of SDW order makes it harder to detect Kitaev physics through inelastic scattering at low energies, we suggest for future experiments, the use of a magnetic field to suppress this order.  As an experimental test for our proposal, we predict a strong suppression of the low-temperature entropy by an external magnetic field well below the Curie-Weiss temperature.
	
	In the Kitaev materials $\alpha-$RuCl$_{3}$ and Na$_{2}$IrO$_{3},$ the coupling $K>0$ but the SDW order is zigzag type, usually stabilized through two distinct physical routes, namely anisotropic interactions $\Gamma, \Gamma',$ \cite{gordon2019theory} or further neighbour Heisenberg interactions \cite{katukuri2014kitaev} $J_2-J_3.$  In the $\Gamma-\Gamma'$ model, increasing $\Gamma$ causes a transition to a non-Kitaev deconfined phase \cite{gordon2019theory} with $\langle W_p \rangle = -0.33,$ that is weakly affected upon inducing zigzag order by turning on $\Gamma'.$  In contrast, in the $J_2-J_3$ model, we have checked (see SM) that Kitaev states are stable inside the PSL, and the fluxes as well as vison gap smoothly decrease upon increasing the competing interactions.  Going beyond Kitaev, our work illustrates persistence of spinonic effects in SDW phases near spin liquids, supporting a similar claim made recently in the context of the cuprates\cite{Sachdevcuprates}.
	
	We end with some comments on the relevance of our work for understanding MBL transitions in disorder free systems. 
	Recent developments have
	\cite{Moessner_disorder_free_localization,Brenes_gauge_invariance_mbl,diamantini2018confinement}
	shown that MBL transitions are also possible in disorder-free systems, with both the (many-body) localized and delocalized phases nonergodic \cite{Moessner_Absence_ergodicity}.
	We find that the level-spacing distribution remains
	Poisson-like in the entire parameter space (see SM), so that our model
	is always non-ergodic, and yet exhibits the above MBL transitions. A key feature of the MBL transition in disordered interacting systems\cite{altshuler1997,mirlin1997}
	is the presence of a mobility edge, which has not been studied in
	the disorder-free context in Refs. \cite{Moessner_disorder_free_localization,Brenes_gauge_invariance_mbl,diamantini2018confinement,Moessner_Absence_ergodicity}.
	Our finite size scaling analysis indicates the presence of two mobility edges separating
	many-body localized, fractal and delocalized phases.
	\begin{acknowledgments}
		We gratefully acknowledge useful discussions with S. Bhattacharjee,
		K. Damle, S. Kundu, T. Senthil, N. Trivedi and V. Vinokur. Special
		thanks to S. Biswas for bringing to our attention the FEAST eigensolver
		scheme. A.K. thanks Pranav S. for help with usage of computational resources. We also acknowledge support of Department of Theoretical Physics,
		TIFR, for computational resources, and V.T. thanks DST for a Swarnajayanti
		grant (No. DST/SJF/PSA- 0212012-13).
	\end{acknowledgments}
	
	\appendix
	
	\section{localization of two-vortex states in Fock space}
	
	\subsection{Overlap of two-vortex Kitaev state with exact eigenstates of $J-K$
		model}
	
\begin{figure}
		\includegraphics[width=1\columnwidth]{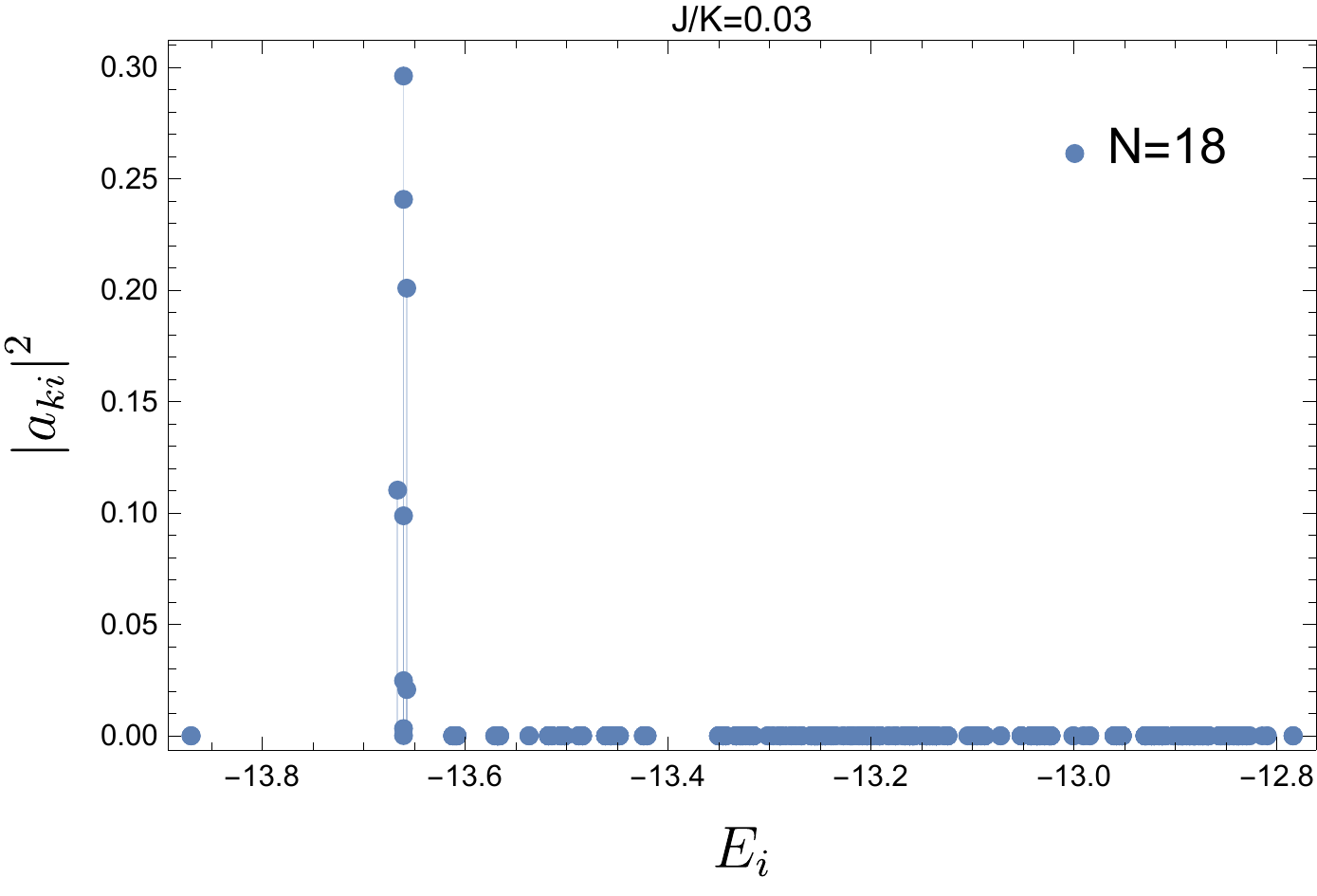}\medskip{}
		
		\includegraphics[width=1\columnwidth]{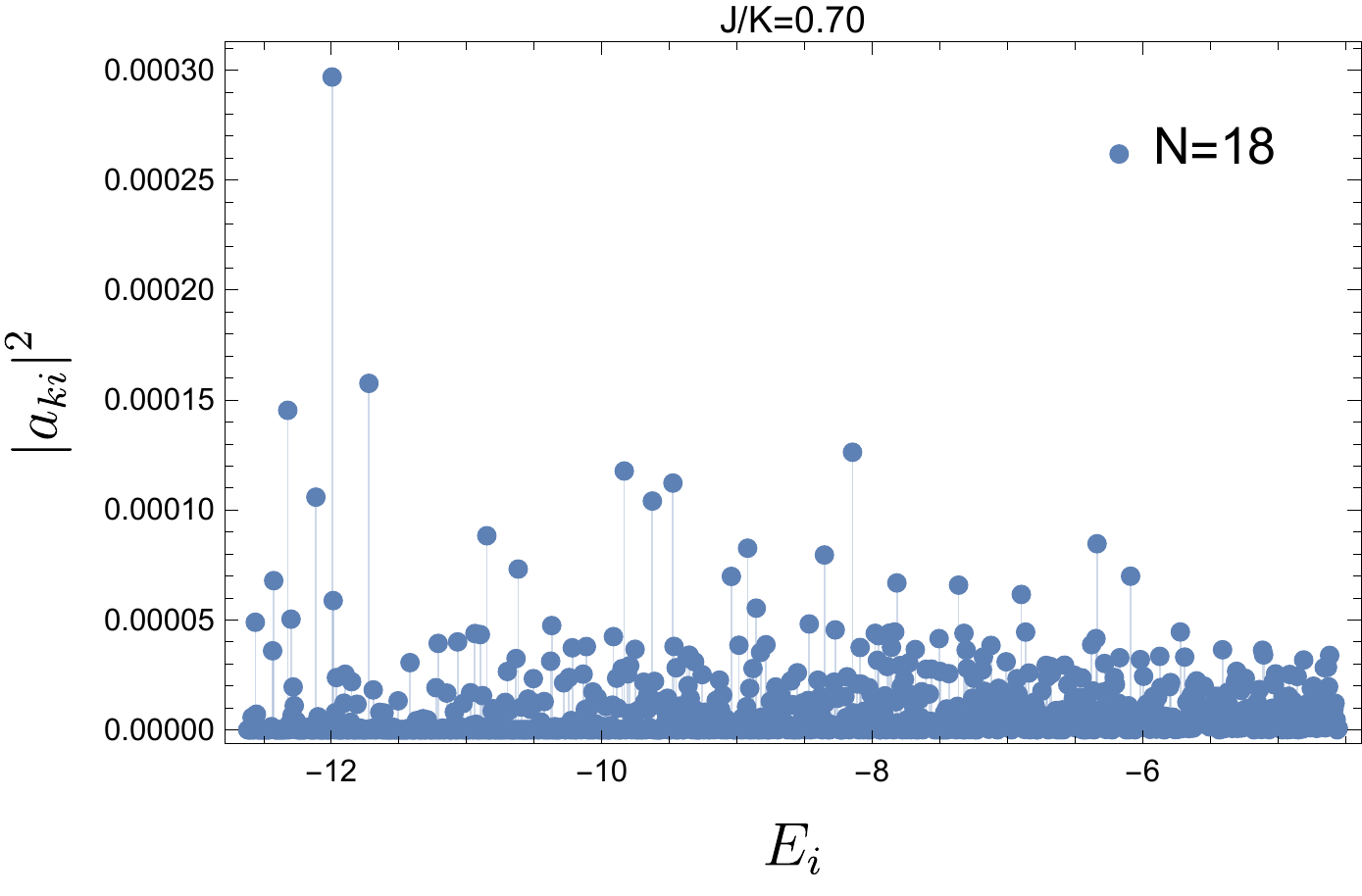}
		
		\caption{\label{fig:overlap-1}Plot showing the squares of the overlap of a
			two-vortex state of the Kitaev model with exact many-body states corresponding
			to two different values of $J/K.$ The exact eigenstates are labeled
			by their energies. For $J/K=0.03,$ the two-vortex state has a small
			support size in the Fock space, and is localized, while for $J/K=0.7,$
			this state is essentially delocalized. Localization and delocalization
			are inferred from the finite size scaling behaviors of the support
			sizes.}
	\end{figure}

	As the Heisenberg perturbation is increased, the Kitaev states begin
	overlapping significantly with an increasing number of exact many-body
	states of the $J-K$ model. Here we show how the support size of a
	two-vortex state, lying close to the ground state of the Kitaev model,
	increases with the ratio $J/K.$ Figure \ref{fig:overlap-1} shows
	a plot of the squares of the overlap, $|a_{ki}|^{2},$ of a two-vortex
	state of the Kitaev model with exact many-body states corresponding
	to two different values of $J/K,$ where $|k\rangle=\sum_{i}a_{ki}|i\rangle$
	is the two-vortex Kitaev state and $\{|i\rangle\}$ are the exact
	eigenstates of the $J-K$ model. For $J/K=0.03,$ the two-vortex state
	has a very small support size in the Fock space, while for $J/K=0.7,$
	the support size is large. Finite size scaling behaviors of the support
	sizes (Figure 1 in main text) tells us about the localization of these
	states. For $J/K=0.03,$ this two-vortex state is localized in Fock
	space while for $J/K=0.7,$ the state is delocalized.
	
	\subsection{Finite size scaling of entanglement entropy}
	
	\begin{figure}
		\includegraphics[width=1\columnwidth]{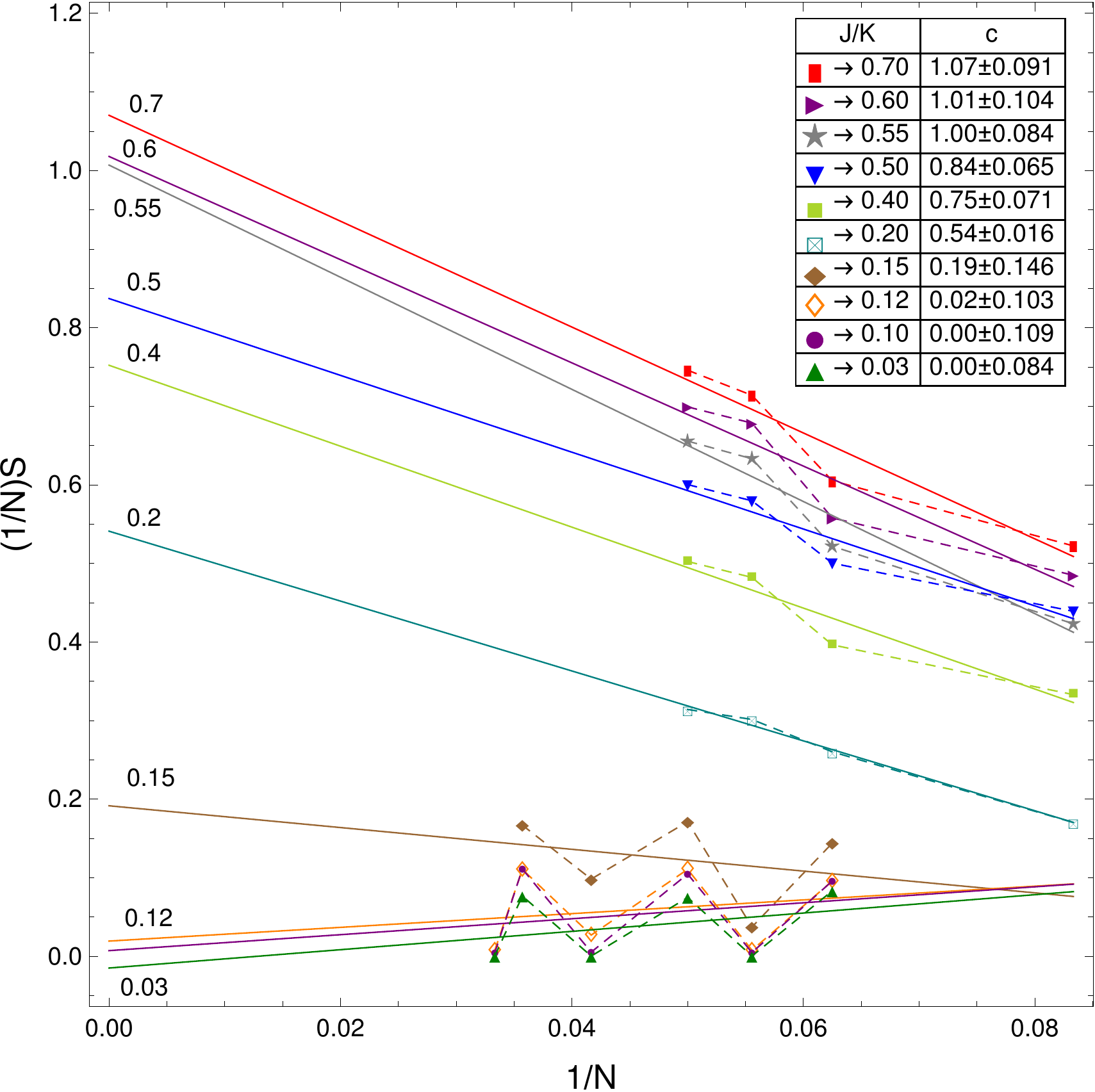}
		
		\caption{\label{fig:entropy-1}Plot describing finite size scaling of the entanglement
			entropy, $S,$ of the lowest two-vortex Kitaev state for different
			values of $J/K.$ The fits are to a volume law, $S/N=c-\frac{1}{N}\ln_{2}(1/f),$
			where $f<1$ and scales slower than exponential. Lines with negative
			(positive) slopes correspond to many-body delocalized (localized)
			phases. The numbers on the solid lines indicate the value of $J/K.$}
	\end{figure}
	
	Apart from the inverse participation ratio, we also analyzed the scaling
	behavior of the entanglement entropy $S_{k},$ of the $k^{{\rm th}}$
	Kitaev state with the exact eigenstates, $|i\rangle,$ of the $J-K$
	model\cite{Khaliullin}: 
	\begin{align}
	S_{k} & =-\sum_{i=1}^{D}|a_{ki}|^{2}\log_{2}|a_{ki}|^{2},\label{eq:shannon-S-1}
	\end{align}
	The entanglement entropy shows a very similar scaling behavior as
	that of the support size, $\xi_{k}.$ In Fig. \ref{fig:entropy-1},
	we show the scaling of the entanglement entropy of the two-flux state
	with $N$ for different values of $J/K.$ In the localized regime,
	$J/K\lesssim0.12,$ the slope of $S_{k}/N$ vs. $1/N$ is positive,
	while it is negative in the fractal and delocalized regimes. In the
	completely delocalized regime, $J/K\gtrsim0.5,$ the entropy increases
	linearly with $N$ with approximately unit slope.
	
	\section{Support sizes of higher Kitaev states}
	
\begin{figure}
		\includegraphics[width=1\columnwidth]{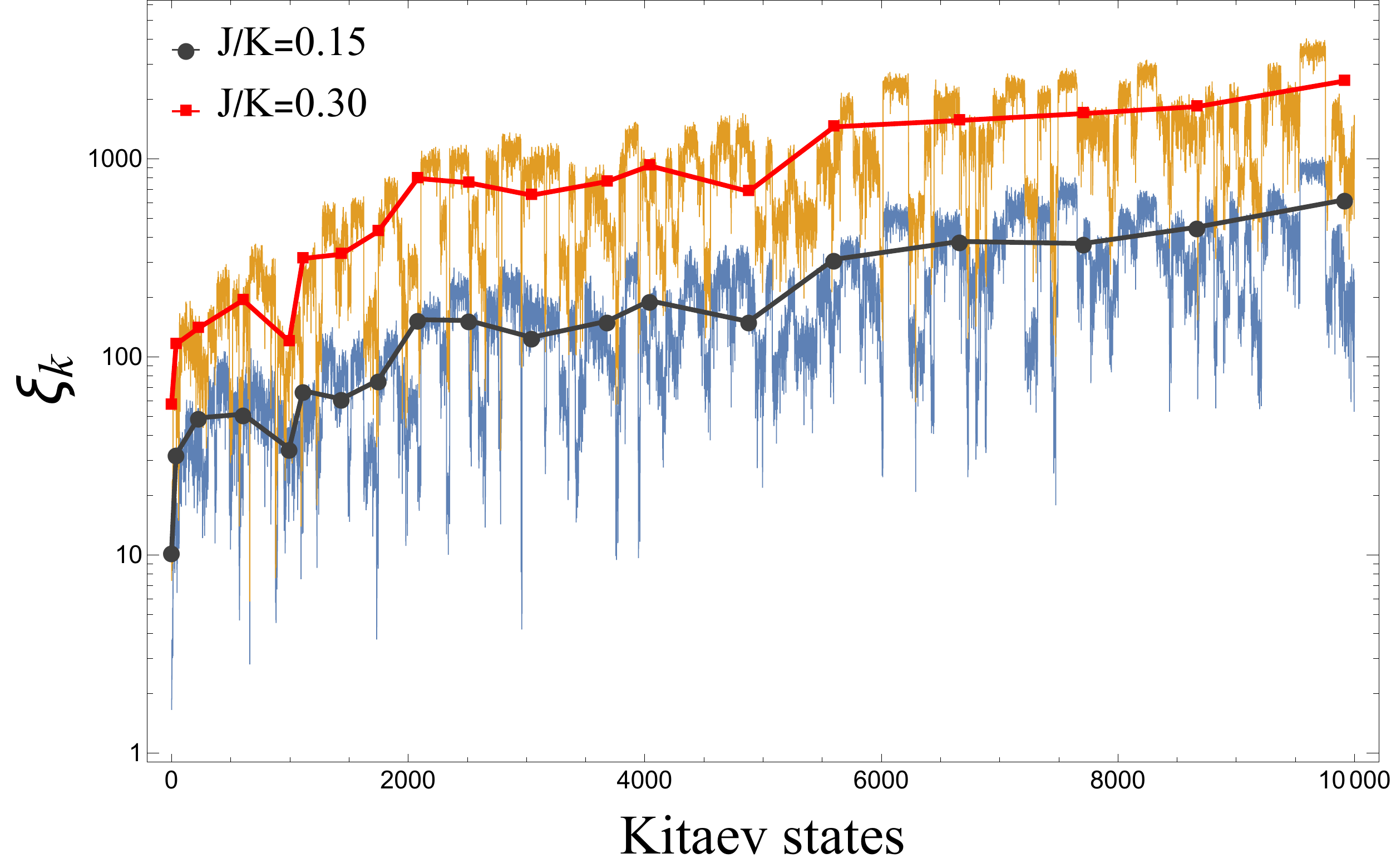}
		
		\caption{\label{fig:support_general-1}Figure showing the support sizes of
			the lowest Kitaev states in the Fock space of exact eigenstates of
			the $J-K$ model, for two different values of $J/K.$ The data is
			for an 18-site cluster, and support sizes of the lowest 12,000 states
			are shown. The flat regions correspond to nearly degenerate Kitaev
			states. Barring some rare exceptions, Kitaev states with comparable
			energies have comparable support sizes. The solid curves represent
			the average support sizes in a fixed small energy interval. Note the
			overall increase of the support size of the excited states with increasing
			$J/K.$ }
	\end{figure}

	Figure \ref{fig:support_general-1} shows that the Kitaev states with
	comparable energies have comparable support sizes. The flat regions
	correspond to nearly degenerate Kitaev states. Some rare states appear
	to have anomalously low support sizes - these form a small fraction
	of the total, and do not affect the average trend of increasing support
	size with energy. This justifies regarding the energy density as an
	appropriate parameter for describing Fock space localization of Kitaev
	states. 

	\section{PHASE DIAGRAM FOR STABILITY OF KITAEV STATES}
	
\begin{figure}
		\includegraphics[width=1\columnwidth]{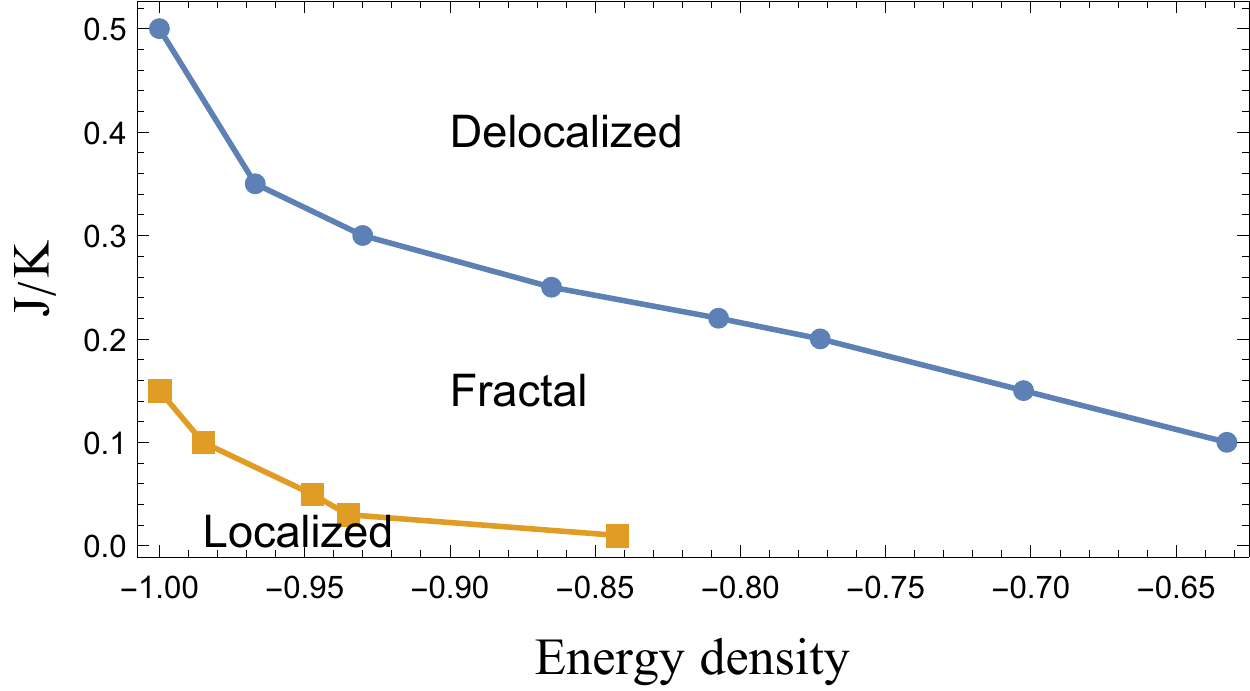}
		
		\caption{\label{fig:phases}Phase diagram showing different finite size scaling
			(up to 24 sites) regimes of Kitaev quasipaticle states as a function
			of $J/K$ and energy density. The PSL phase corresponds to $J/K\gtrsim0.12.$
			Kitaev states are not localized in the PSL phase but are nevertheless
			stable over a significant low energy range where they show fractal
			scaling. States corresponding to high energy densities are delocalized
			and thus not Kitaev-like.}
	\end{figure}

	Figure \ref{fig:phases} shows the phase diagram of Kitaev quasiparticle
	stability based on the finite size scaling behavior in the $J/K$
	vs. energy density plane. Low energy Kitaev states are clearly more
	robust against delocalization by a Heisenberg perturbation compared
	to the higher energy states. Near the middle of the excitation spectrum,
	Kitaev states get delocalized even by small perturbations. We also
	found the highest positive energy excited states show a similar behavior
	as those close to the ground state (not shown in the Figure).
	
	\section{Energy level statistics}
	
\begin{figure}
		\includegraphics[width=1\columnwidth]{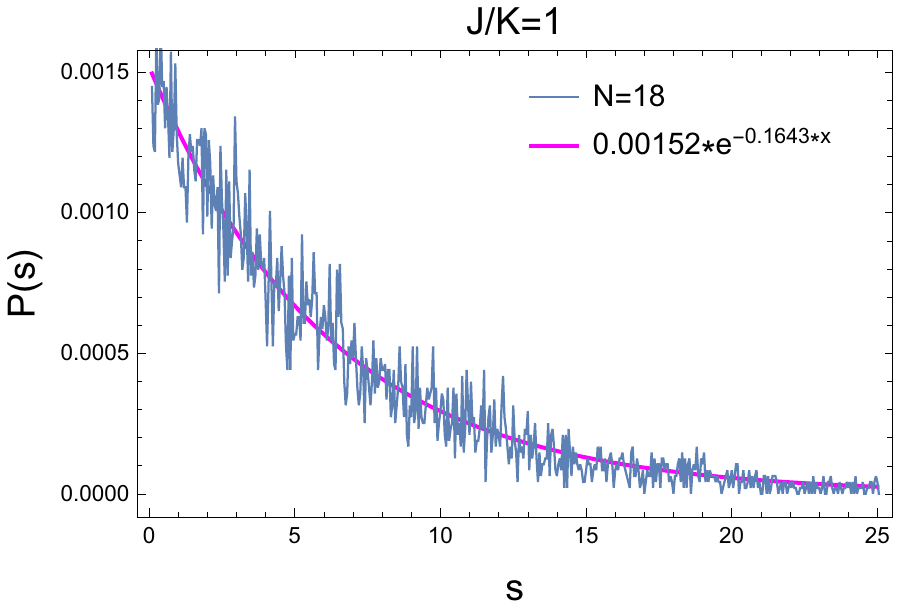}
		
		\caption{\label{fig:poisson-1}Plot showing the distribution $P(s)$ of energy
			level spacings $s$ (in units of the mean level spacing $\delta$)
			obtained for an $N=18$ cluster, for $J/K=1$ that corresponds to
			the fully delocalized regime for the Kitaev states. The fit is to
			a Poisson distribution, $P(s)=A\exp[-bs],$ where $A=$0.00152 and
			$b=0.16432$. The Poissonian level statistics is seen throughout the
			parameter space corresponding to localized, fractal and delocalized
			regimes. The Fock space transitions to fractal and completely delocalized
			phases occur completely within the nonergodic regime in our disorder-free
			model.}
	\end{figure}

	Figure \ref{fig:poisson-1} shows the distribution $P(s)$ of energy
	level spacings $s,$ measured in units of the mean level spacing $\delta,$
	for an $18$-site cluster, for $J/K=1$ corresponding to the fully
	delocalized regime for the Kitaev states. The distribution fits well
	to a Poisson law and not Wigner-Dyson, showing that the model is nonergodic
	in this fully delocalized regime. We found that the level statistics
	remains Poisson like throught the localized, fractal and delocalized
	regimes, supporting the view that in disorder-free models, many-body
	localization transitions are possible without an accompanying ergodic-nonergodic
	transition. For $s=0$ (not shown in the plot), $P(s)$ takes a rather
	large value $\sim0.82$ owing to the large number of degenerate or
	nearly degenerate states near the middle of the spectrum.
	
	\section{Specific heat anomalies}
	
\begin{figure}
		\includegraphics[width=1\columnwidth]{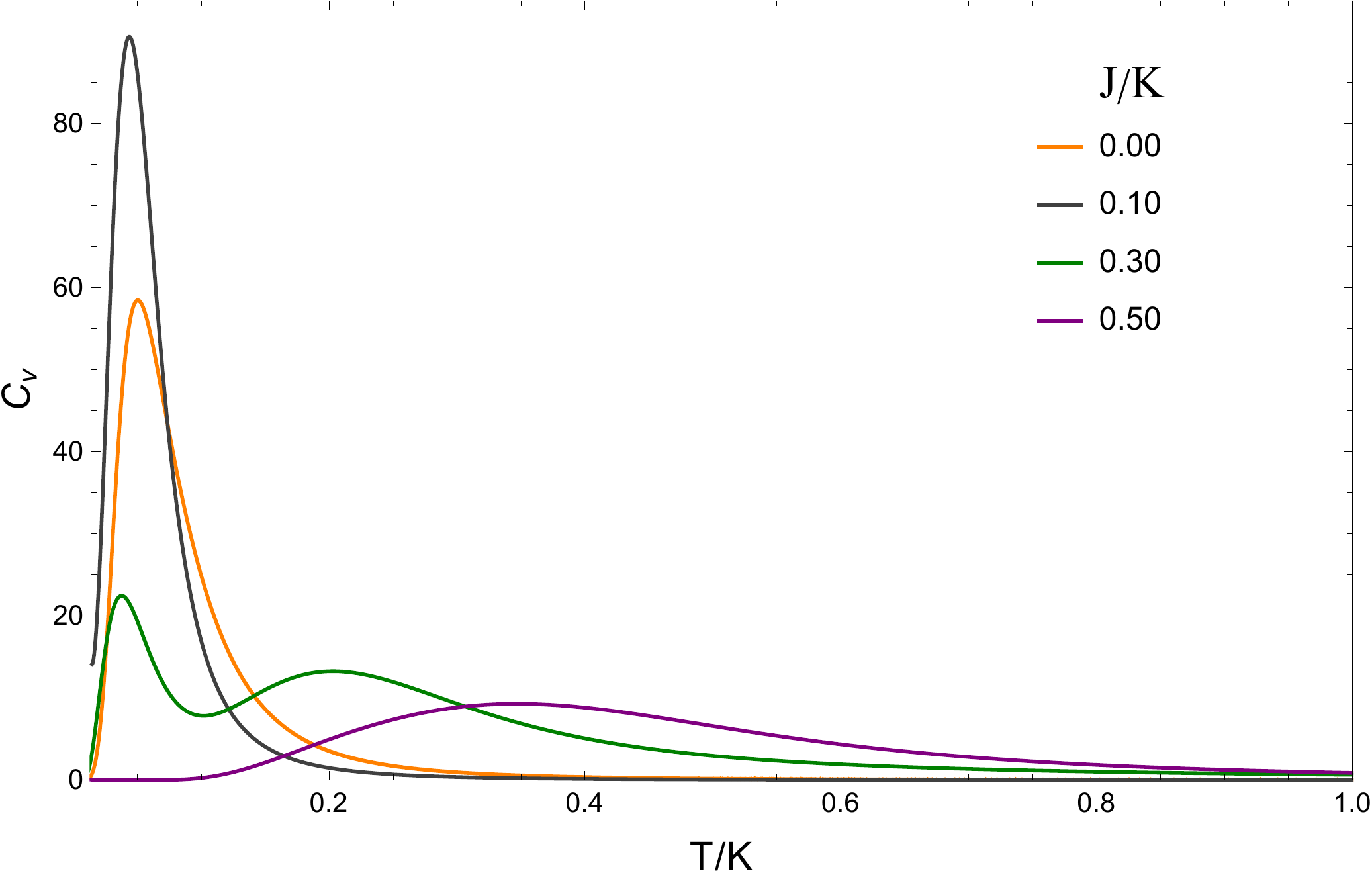}
		
		\caption{\label{fig:specific_heat}Plot showing specific heat vs $T/K$ for
			a 24 site cluster in the vicinity of the KSL-PSL boundary at $J/K\approx0.13.$
			A single peak is visible in the KSL phase representing the vison gap,
			while in the PSL phase, an additional peak appears at higher temperatures
			due to appearance of SDW order. The vison peak persists in the SDW
			phase up to the peak of the stripy phase.}
	\end{figure}

	Figure \ref{fig:specific_heat} shows the specific heat ($c_{v}$)
	vs. $T/K$ phase diagram for a 24-site cluster for different values
	of $J/K$ straddling the KSL and PSL phases. In the KSL phase, there
	is only one peak (beginning from around $0.05K$ for the pure Kitaev
	model) representing the vison gap. This vison peak keeps shifting
	slowly towards lower tempertaures as we increase $J/K$. As we enter
	the PSL phase at $J/K\approx0.13,$ another peak appears at higher
	temperature associated with onset of stripy SDW order. As we increase
	$J,$ this stripy SDW peak keeps shifting towards higher temperatures,
	while the vison peak feature drifts towards lower tempertures vanishing
	only at $J/K=0.5$ where stripy SDW order peaks.
	
	\section{Vison Gap in KSL and PSL phases}
	As a further check on the interpretation of the specific heat peak at lower temperatures, we have also directly estimated vison gap for different values of $J/K$ for a 24-site cluster. Following the prescription in Kitaev's original paper, the vison gap is estimates as half the energy of the lowest two-flux excitation, with the vortices as far apart as possible. The wavefunctions used are respectively the Kitaev ground state and the lowest two-flux state. Table~\ref{tab:visongap} shows the calculated vison gaps for $J/K$ values straddling the KSL/PSL phase boundary ($J/K=0.125$). These values are in good agreement with the peak positions in the specific heat.
	
	\begin{table}
		\begin{tabular}{ |c|c|c|} 
			\hline
			&$J/K$ & Vison gap ($\times K$) \\
			\hline
			&0 & 0.068 \\ 
			&0.10 & 0.060 \\ 
			&0.20 & 0.053 \\ 
			&0.30 & 0.045 \\
			&0.40 & 0.04 \\
			\hline
		\end{tabular}
		\caption{\label{tab:visongap} Calculated values of the vison gap from the energy of the lowest two-flux state. The calculated values are in good agreement with the low-temperature peak positions of the specific heat, providing additional support that these peaks are associated with a vison gap.}
	\end{table}
	
	\section{$J_{2}-J_{3}$ Model}
	
\begin{figure}
		\includegraphics[width=1\columnwidth]{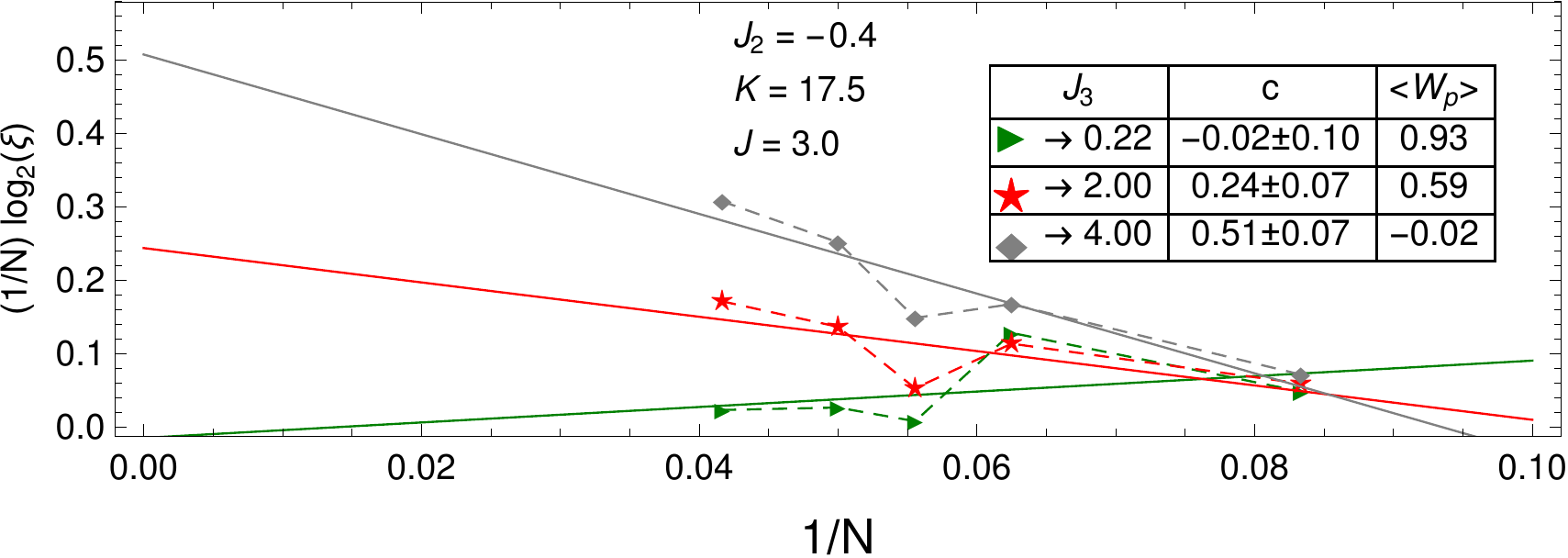}
		\caption{\label{fig:J3-scaling}Finite size scaling for the pure Kitaev ground
			state in the Fock space of the exact eigenstates of the $J_{2}-J_{3}$
			model for three different values of $J_{3}$ spanning the KSL to zigzag
			SDW transition. The parameter values are chosen from Ref. \cite{katukuri2014kitaev}.
			$J_{3}=0.22$ is within the KSL phase, $J_{3}=2$ is on the zigzag
			SDW side near the KSL-SDW boundary, while $J_{3}=4$ is deeper in
			the SDW phase. The table in the inset shows the calculated values
			of the exponent $c$ and the plaquette flux $W_{p}.$ The Kitaev ground
			state shows fractal scaling for the chosen values of $J_{3}$ in the
			SDW phase, and is localized for $J_{3}=0.22.$ The value of $W_{p}$
			does not vanish upon entering the SDW phase but shows a smooth declining
			trend moving into the SDW phase. }
	\end{figure}

	We analyze the following $J_{2}-J_{3}$ model\cite{katukuri2014kitaev,torque_responce,Kitaev-Heisenbeeg-J2_J3_PhysRev}
	for the Kitaev materials,
	
	\begin{align}
	H & =-K\sum_{\langle ij\rangle,\gamma}\sigma_{i}^{\gamma}\sigma_{j}^{\gamma}+J\sum_{\langle ij\rangle}\boldsymbol{\sigma}_{i}\cdot\boldsymbol{\sigma}_{j}\label{eq:model}\\
	& +J_{2}\sum_{\langle\langle ij\rangle\rangle}\boldsymbol{\sigma}_{i}\cdot\boldsymbol{\sigma}_{j}+J_{3}\sum_{\langle\langle\langle ij\rangle\rangle\rangle}\boldsymbol{\sigma}_{i}\cdot\boldsymbol{\sigma}_{j},\nonumber 
	\end{align}
	where $\langle\rangle$,$\langle\langle\rangle\rangle$,$\langle\langle\langle\rangle\rangle\rangle$
	represents first, second and third nearest nearest neighbours respectively.
	We study the localization of the pure Kitaev ground state in the Fock
	space of exact eigenstates of the full $J_{2}-J_{3}$ model by tuning
	the $J_{3}$ parameter (with fixed $K,\,J,\,J_{2}$) spanning the
	KSL to zigzag SDW regimes. The corresponding plaquette fluxes are
	also evaluated. For concreteness, three representative different values
	of $J_{3}$ are chosen from Ref. \cite{katukuri2014kitaev} - $J_{3}=0.22$
	(KSL phase), $J_{3}=2$ (zigzag SDW phase, near KSL-SDW boundary)
	and $J_{3}=4$ (deeper in SDW phase). Figure \ref{fig:J3-scaling}
	shows the finite size scaling analysis for the support size of the
	Kitaev ground state. The inset shows the calculated values of the
	scaling exponent $c,$ as well as the plaquette fluxes. For $J_{3}=0.22,$
	the Kitaev ground state is evidently localized (positive slope with
	$c\approx0$), and plaquette fluxes are close to $W_{p}=1$ expected
	for the pure KSL limit. At the two higher values of $J_{3},$ the
	support size of the Kitaev ground state shows fractal scaling, i.e.,
	the Kitaev states are still stable even in this zigzag SDW phase.
	Note the exponent $c$ increases from $0.24$ for $J_{3}=2$ to $0.51$
	for $J_{3}=4,$ representing increasing delocalization away from the
	KSL. The corresponding values of $W_{p}$ also decline sharply. The
	stability of Kitaev states in the SDW phase of the $J_{2}-J_{3}$
	model distinguishes it from the alternate $\Gamma-\Gamma'$ model\cite{gordon2019theory}
	also studied in the context of Kitaev materials. In the $\Gamma-\Gamma'$
	model, the zigzag phase is known to be deconfined but not Kitaev-like\cite{gordon2019theory}\textbf{
	}. The behavior of the $J_{2}-J_{3}$ model is similar to the simple
	$J-K$ model studied in our paper.
	
\begin{thebibliography}{57}%
	\makeatletter
	\providecommand \@ifxundefined [1]{%
		\@ifx{#1\undefined}
	}%
	\providecommand \@ifnum [1]{%
		\ifnum #1\expandafter \@firstoftwo
		\else \expandafter \@secondoftwo
		\fi
	}%
	\providecommand \@ifx [1]{%
		\ifx #1\expandafter \@firstoftwo
		\else \expandafter \@secondoftwo
		\fi
	}%
	\providecommand \natexlab [1]{#1}%
	\providecommand \enquote  [1]{``#1''}%
	\providecommand \bibnamefont  [1]{#1}%
	\providecommand \bibfnamefont [1]{#1}%
	\providecommand \citenamefont [1]{#1}%
	\providecommand \href@noop [0]{\@secondoftwo}%
	\providecommand \href [0]{\begingroup \@sanitize@url \@href}%
	\providecommand \@href[1]{\@@startlink{#1}\@@href}%
	\providecommand \@@href[1]{\endgroup#1\@@endlink}%
	\providecommand \@sanitize@url [0]{\catcode `\\12\catcode `\$12\catcode
		`\&12\catcode `\#12\catcode `\^12\catcode `\_12\catcode `\%12\relax}%
	\providecommand \@@startlink[1]{}%
	\providecommand \@@endlink[0]{}%
	\providecommand \url  [0]{\begingroup\@sanitize@url \@url }%
	\providecommand \@url [1]{\endgroup\@href {#1}{\urlprefix }}%
	\providecommand \urlprefix  [0]{URL }%
	\providecommand \Eprint [0]{\href }%
	\providecommand \doibase [0]{http://dx.doi.org/}%
	\providecommand \selectlanguage [0]{\@gobble}%
	\providecommand \bibinfo  [0]{\@secondoftwo}%
	\providecommand \bibfield  [0]{\@secondoftwo}%
	\providecommand \translation [1]{[#1]}%
	\providecommand \BibitemOpen [0]{}%
	\providecommand \bibitemStop [0]{}%
	\providecommand \bibitemNoStop [0]{.\EOS\space}%
	\providecommand \EOS [0]{\spacefactor3000\relax}%
	\providecommand \BibitemShut  [1]{\csname bibitem#1\endcsname}%
	\let\auto@bib@innerbib\@empty
	\bibitem [{\citenamefont {Kitaev}(2006)}]{Kitaev:2006lla}%
	\BibitemOpen
	\bibfield  {author} {\bibinfo {author} {\bibfnamefont {A.}~\bibnamefont
			{Kitaev}},\ }\href {\doibase 10.1016/j.aop.2005.10.005} {\bibfield  {journal}
		{\bibinfo  {journal} {Annals Phys.}\ }\textbf {\bibinfo {volume} {321}},\
		\bibinfo {pages} {2} (\bibinfo {year} {2006})}\BibitemShut {NoStop}%
	\bibitem [{\citenamefont {Gohlke}\ \emph {et~al.}(2017)\citenamefont {Gohlke},
		\citenamefont {Verresen}, \citenamefont {Moessner},\ and\ \citenamefont
		{Pollmann}}]{gohlke2017dynamics}%
	\BibitemOpen
	\bibfield  {author} {\bibinfo {author} {\bibfnamefont {M.}~\bibnamefont
			{Gohlke}}, \bibinfo {author} {\bibfnamefont {R.}~\bibnamefont {Verresen}},
		\bibinfo {author} {\bibfnamefont {R.}~\bibnamefont {Moessner}}, \ and\
		\bibinfo {author} {\bibfnamefont {F.}~\bibnamefont {Pollmann}},\ }\href
	{\doibase 10.1103/PhysRevLett.119.157203} {\bibfield  {journal} {\bibinfo
			{journal} {Phys. Rev. Lett.}\ }\textbf {\bibinfo {volume} {119}},\ \bibinfo
		{pages} {157203} (\bibinfo {year} {2017})}\BibitemShut {NoStop}%
	\bibitem [{\citenamefont {Winter}\ \emph {et~al.}(2017)\citenamefont {Winter},
		\citenamefont {Riedl}, \citenamefont {Maksimov}, \citenamefont {Chernyshev},
		\citenamefont {Honecker},\ and\ \citenamefont
		{Valent{\'\i}}}]{winter2017breakdown}%
	\BibitemOpen
	\bibfield  {author} {\bibinfo {author} {\bibfnamefont {S.~M.}\ \bibnamefont
			{Winter}}, \bibinfo {author} {\bibfnamefont {K.}~\bibnamefont {Riedl}},
		\bibinfo {author} {\bibfnamefont {P.~A.}\ \bibnamefont {Maksimov}}, \bibinfo
		{author} {\bibfnamefont {A.~L.}\ \bibnamefont {Chernyshev}}, \bibinfo
		{author} {\bibfnamefont {A.}~\bibnamefont {Honecker}}, \ and\ \bibinfo
		{author} {\bibfnamefont {R.}~\bibnamefont {Valent{\'\i}}},\ }\href@noop {}
	{\bibfield  {journal} {\bibinfo  {journal} {Nature Communications}\ }\textbf
		{\bibinfo {volume} {8}},\ \bibinfo {pages} {1152} (\bibinfo {year}
		{2017})}\BibitemShut {NoStop}%
	\bibitem [{\citenamefont {Gordon}\ \emph {et~al.}(2019)\citenamefont {Gordon},
		\citenamefont {Catuneanu}, \citenamefont {S{\o}rensen},\ and\ \citenamefont
		{Kee}}]{gordon2019theory}%
	\BibitemOpen
	\bibfield  {author} {\bibinfo {author} {\bibfnamefont {J.~S.}\ \bibnamefont
			{Gordon}}, \bibinfo {author} {\bibfnamefont {A.}~\bibnamefont {Catuneanu}},
		\bibinfo {author} {\bibfnamefont {E.~S.}\ \bibnamefont {S{\o}rensen}}, \ and\
		\bibinfo {author} {\bibfnamefont {H.-Y.}\ \bibnamefont {Kee}},\ }\href@noop
	{} {\bibfield  {journal} {\bibinfo  {journal} {Nature Communications}\
		}\textbf {\bibinfo {volume} {10}},\ \bibinfo {pages} {2470} (\bibinfo {year}
		{2019})}\BibitemShut {NoStop}%
	\bibitem [{\citenamefont {Chaloupka}\ \emph {et~al.}(2010)\citenamefont
		{Chaloupka}, \citenamefont {Jackeli},\ and\ \citenamefont
		{Khaliullin}}]{Khaliullin}%
	\BibitemOpen
	\bibfield  {author} {\bibinfo {author} {\bibfnamefont {J.}~\bibnamefont
			{Chaloupka}}, \bibinfo {author} {\bibfnamefont {G.}~\bibnamefont {Jackeli}},
		\ and\ \bibinfo {author} {\bibfnamefont {G.}~\bibnamefont {Khaliullin}},\
	}\href {\doibase 10.1103/PhysRevLett.105.027204} {\bibfield  {journal}
		{\bibinfo  {journal} {Phys. Rev. Lett.}\ }\textbf {\bibinfo {volume} {105}},\
		\bibinfo {pages} {027204} (\bibinfo {year} {2010})}\BibitemShut {NoStop}%
	\bibitem [{\citenamefont {Kimchi}\ and\ \citenamefont
		{You}(2011)}]{Kitaev-Heisenbeeg-J2_J3_PhysRev}%
	\BibitemOpen
	\bibfield  {author} {\bibinfo {author} {\bibfnamefont {I.}~\bibnamefont
			{Kimchi}}\ and\ \bibinfo {author} {\bibfnamefont {Y.-Z.}\ \bibnamefont
			{You}},\ }\href {\doibase 10.1103/PhysRevB.84.180407} {\bibfield  {journal}
		{\bibinfo  {journal} {Phys. Rev. B}\ }\textbf {\bibinfo {volume} {84}},\
		\bibinfo {pages} {180407} (\bibinfo {year} {2011})}\BibitemShut {NoStop}%
	\bibitem [{\citenamefont {Katukuri}\ \emph {et~al.}(2014)\citenamefont
		{Katukuri}, \citenamefont {Nishimoto}, \citenamefont {Yushankhai},
		\citenamefont {Stoyanova}, \citenamefont {Kandpal}, \citenamefont {Choi},
		\citenamefont {Coldea}, \citenamefont {Rousochatzakis}, \citenamefont
		{Hozoi},\ and\ \citenamefont {Van Den~Brink}}]{katukuri2014kitaev}%
	\BibitemOpen
	\bibfield  {author} {\bibinfo {author} {\bibfnamefont {V.~M.}\ \bibnamefont
			{Katukuri}}, \bibinfo {author} {\bibfnamefont {S.}~\bibnamefont {Nishimoto}},
		\bibinfo {author} {\bibfnamefont {V.}~\bibnamefont {Yushankhai}}, \bibinfo
		{author} {\bibfnamefont {A.}~\bibnamefont {Stoyanova}}, \bibinfo {author}
		{\bibfnamefont {H.}~\bibnamefont {Kandpal}}, \bibinfo {author} {\bibfnamefont
			{S.}~\bibnamefont {Choi}}, \bibinfo {author} {\bibfnamefont {R.}~\bibnamefont
			{Coldea}}, \bibinfo {author} {\bibfnamefont {I.}~\bibnamefont
			{Rousochatzakis}}, \bibinfo {author} {\bibfnamefont {L.}~\bibnamefont
			{Hozoi}}, \ and\ \bibinfo {author} {\bibfnamefont {J.}~\bibnamefont {Van
				Den~Brink}},\ }\href@noop {} {\bibfield  {journal} {\bibinfo  {journal} {New
				Journal of Physics}\ }\textbf {\bibinfo {volume} {16}},\ \bibinfo {pages}
		{013056} (\bibinfo {year} {2014})}\BibitemShut {NoStop}%
	\bibitem [{\citenamefont {Sizyuk}\ \emph {et~al.}(2014)\citenamefont {Sizyuk},
		\citenamefont {Price}, \citenamefont {W\"olfle},\ and\ \citenamefont
		{Perkins}}]{Sizyuk_Phys_iridates}%
	\BibitemOpen
	\bibfield  {author} {\bibinfo {author} {\bibfnamefont {Y.}~\bibnamefont
			{Sizyuk}}, \bibinfo {author} {\bibfnamefont {C.}~\bibnamefont {Price}},
		\bibinfo {author} {\bibfnamefont {P.}~\bibnamefont {W\"olfle}}, \ and\
		\bibinfo {author} {\bibfnamefont {N.~B.}\ \bibnamefont {Perkins}},\ }\href
	{\doibase 10.1103/PhysRevB.90.155126} {\bibfield  {journal} {\bibinfo
			{journal} {Phys. Rev. B}\ }\textbf {\bibinfo {volume} {90}},\ \bibinfo
		{pages} {155126} (\bibinfo {year} {2014})}\BibitemShut {NoStop}%
	\bibitem [{\citenamefont {Rau}\ \emph {et~al.}(2016)\citenamefont {Rau},
		\citenamefont {Lee},\ and\ \citenamefont
		{Kee}}]{annurev-conmatphys-031115-011319}%
	\BibitemOpen
	\bibfield  {author} {\bibinfo {author} {\bibfnamefont {J.~G.}\ \bibnamefont
			{Rau}}, \bibinfo {author} {\bibfnamefont {E.~K.-H.}\ \bibnamefont {Lee}}, \
		and\ \bibinfo {author} {\bibfnamefont {H.-Y.}\ \bibnamefont {Kee}},\ }\href
	{\doibase 10.1146/annurev-conmatphys-031115-011319} {\bibfield  {journal}
		{\bibinfo  {journal} {Annual Review of Condensed Matter Physics}\ }\textbf
		{\bibinfo {volume} {7}},\ \bibinfo {pages} {195} (\bibinfo {year}
		{2016})}\BibitemShut {NoStop}%
	\bibitem [{\citenamefont {Yamaji}\ \emph {et~al.}(2014)\citenamefont {Yamaji},
		\citenamefont {Nomura}, \citenamefont {Kurita}, \citenamefont {Arita},\ and\
		\citenamefont {Imada}}]{Youhei_iridates_PhysRevLett}%
	\BibitemOpen
	\bibfield  {author} {\bibinfo {author} {\bibfnamefont {Y.}~\bibnamefont
			{Yamaji}}, \bibinfo {author} {\bibfnamefont {Y.}~\bibnamefont {Nomura}},
		\bibinfo {author} {\bibfnamefont {M.}~\bibnamefont {Kurita}}, \bibinfo
		{author} {\bibfnamefont {R.}~\bibnamefont {Arita}}, \ and\ \bibinfo {author}
		{\bibfnamefont {M.}~\bibnamefont {Imada}},\ }\href {\doibase
		10.1103/PhysRevLett.113.107201} {\bibfield  {journal} {\bibinfo  {journal}
			{Phys. Rev. Lett.}\ }\textbf {\bibinfo {volume} {113}},\ \bibinfo {pages}
		{107201} (\bibinfo {year} {2014})}\BibitemShut {NoStop}%
	\bibitem [{\citenamefont {Bhattacharjee}\ \emph {et~al.}(2012)\citenamefont
		{Bhattacharjee}, \citenamefont {Lee},\ and\ \citenamefont
		{Kim}}]{bhattacharjee2012spin}%
	\BibitemOpen
	\bibfield  {author} {\bibinfo {author} {\bibfnamefont {S.}~\bibnamefont
			{Bhattacharjee}}, \bibinfo {author} {\bibfnamefont {S.-S.}\ \bibnamefont
			{Lee}}, \ and\ \bibinfo {author} {\bibfnamefont {Y.~B.}\ \bibnamefont
			{Kim}},\ }\href@noop {} {\bibfield  {journal} {\bibinfo  {journal} {New
				Journal of Physics}\ }\textbf {\bibinfo {volume} {14}},\ \bibinfo {pages}
		{073015} (\bibinfo {year} {2012})}\BibitemShut {NoStop}%
	\bibitem [{\citenamefont {Hu}\ \emph {et~al.}(2015)\citenamefont {Hu},
		\citenamefont {Wang},\ and\ \citenamefont
		{Feng}}]{Hu_Iridates_PhysRevLett.115.167204}%
	\BibitemOpen
	\bibfield  {author} {\bibinfo {author} {\bibfnamefont {K.}~\bibnamefont
			{Hu}}, \bibinfo {author} {\bibfnamefont {F.}~\bibnamefont {Wang}}, \ and\
		\bibinfo {author} {\bibfnamefont {J.}~\bibnamefont {Feng}},\ }\href {\doibase
		10.1103/PhysRevLett.115.167204} {\bibfield  {journal} {\bibinfo  {journal}
			{Phys. Rev. Lett.}\ }\textbf {\bibinfo {volume} {115}},\ \bibinfo {pages}
		{167204} (\bibinfo {year} {2015})}\BibitemShut {NoStop}%
	\bibitem [{\citenamefont {Shitade}\ \emph {et~al.}(2009)\citenamefont
		{Shitade}, \citenamefont {Katsura}, \citenamefont
		{Kune\ifmmode~\check{s}\else \v{s}\fi{}}, \citenamefont {Qi}, \citenamefont
		{Zhang},\ and\ \citenamefont
		{Nagaosa}}]{quantum_hall_shitade_PhysRevLett.102.256403}%
	\BibitemOpen
	\bibfield  {author} {\bibinfo {author} {\bibfnamefont {A.}~\bibnamefont
			{Shitade}}, \bibinfo {author} {\bibfnamefont {H.}~\bibnamefont {Katsura}},
		\bibinfo {author} {\bibfnamefont {J.}~\bibnamefont
			{Kune\ifmmode~\check{s}\else \v{s}\fi{}}}, \bibinfo {author} {\bibfnamefont
			{X.-L.}\ \bibnamefont {Qi}}, \bibinfo {author} {\bibfnamefont {S.-C.}\
			\bibnamefont {Zhang}}, \ and\ \bibinfo {author} {\bibfnamefont
			{N.}~\bibnamefont {Nagaosa}},\ }\href {\doibase
		10.1103/PhysRevLett.102.256403} {\bibfield  {journal} {\bibinfo  {journal}
			{Phys. Rev. Lett.}\ }\textbf {\bibinfo {volume} {102}},\ \bibinfo {pages}
		{256403} (\bibinfo {year} {2009})}\BibitemShut {NoStop}%
	\bibitem [{\citenamefont {Chaloupka}\ and\ \citenamefont
		{Khaliullin}(2015)}]{Khaliullin_iridates_PhysRevB.92.024413}%
	\BibitemOpen
	\bibfield  {author} {\bibinfo {author} {\bibfnamefont {J.~c.~v.}\
			\bibnamefont {Chaloupka}}\ and\ \bibinfo {author} {\bibfnamefont
			{G.}~\bibnamefont {Khaliullin}},\ }\href {\doibase
		10.1103/PhysRevB.92.024413} {\bibfield  {journal} {\bibinfo  {journal} {Phys.
				Rev. B}\ }\textbf {\bibinfo {volume} {92}},\ \bibinfo {pages} {024413}
		(\bibinfo {year} {2015})}\BibitemShut {NoStop}%
	\bibitem [{\citenamefont {Takayama}\ \emph {et~al.}(2015)\citenamefont
		{Takayama}, \citenamefont {Kato}, \citenamefont {Dinnebier}, \citenamefont
		{Nuss}, \citenamefont {Kono}, \citenamefont {Veiga}, \citenamefont {Fabbris},
		\citenamefont {Haskel},\ and\ \citenamefont
		{Takagi}}]{Takayama_iridates_PhysRevLett.114.077202}%
	\BibitemOpen
	\bibfield  {author} {\bibinfo {author} {\bibfnamefont {T.}~\bibnamefont
			{Takayama}}, \bibinfo {author} {\bibfnamefont {A.}~\bibnamefont {Kato}},
		\bibinfo {author} {\bibfnamefont {R.}~\bibnamefont {Dinnebier}}, \bibinfo
		{author} {\bibfnamefont {J.}~\bibnamefont {Nuss}}, \bibinfo {author}
		{\bibfnamefont {H.}~\bibnamefont {Kono}}, \bibinfo {author} {\bibfnamefont
			{L.~S.~I.}\ \bibnamefont {Veiga}}, \bibinfo {author} {\bibfnamefont
			{G.}~\bibnamefont {Fabbris}}, \bibinfo {author} {\bibfnamefont
			{D.}~\bibnamefont {Haskel}}, \ and\ \bibinfo {author} {\bibfnamefont
			{H.}~\bibnamefont {Takagi}},\ }\href {\doibase
		10.1103/PhysRevLett.114.077202} {\bibfield  {journal} {\bibinfo  {journal}
			{Phys. Rev. Lett.}\ }\textbf {\bibinfo {volume} {114}},\ \bibinfo {pages}
		{077202} (\bibinfo {year} {2015})}\BibitemShut {NoStop}%
	\bibitem [{\citenamefont {Kim}\ \emph {et~al.}(2014)\citenamefont {Kim},
		\citenamefont {Lee},\ and\ \citenamefont {Cho}}]{kim2014antiferromagnetic}%
	\BibitemOpen
	\bibfield  {author} {\bibinfo {author} {\bibfnamefont {H.-J.}\ \bibnamefont
			{Kim}}, \bibinfo {author} {\bibfnamefont {J.-H.}\ \bibnamefont {Lee}}, \ and\
		\bibinfo {author} {\bibfnamefont {J.-H.}\ \bibnamefont {Cho}},\ }\href@noop
	{} {\bibfield  {journal} {\bibinfo  {journal} {Scientific Reports}\ }\textbf
		{\bibinfo {volume} {4}},\ \bibinfo {pages} {5253} (\bibinfo {year}
		{2014})}\BibitemShut {NoStop}%
	\bibitem [{\citenamefont {Chaloupka}\ \emph {et~al.}(2013)\citenamefont
		{Chaloupka}, \citenamefont {Jackeli},\ and\ \citenamefont
		{Khaliullin}}]{Zigzag_khaliullin_PhysRevLett.110.097204}%
	\BibitemOpen
	\bibfield  {author} {\bibinfo {author} {\bibfnamefont {J.~c.~v.}\
			\bibnamefont {Chaloupka}}, \bibinfo {author} {\bibfnamefont {G.}~\bibnamefont
			{Jackeli}}, \ and\ \bibinfo {author} {\bibfnamefont {G.}~\bibnamefont
			{Khaliullin}},\ }\href {\doibase 10.1103/PhysRevLett.110.097204} {\bibfield
		{journal} {\bibinfo  {journal} {Phys. Rev. Lett.}\ }\textbf {\bibinfo
			{volume} {110}},\ \bibinfo {pages} {097204} (\bibinfo {year}
		{2013})}\BibitemShut {NoStop}%
	\bibitem [{\citenamefont {Liu}\ and\ \citenamefont
		{Normand}(2018)}]{Liu2018Dirac}%
	\BibitemOpen
	\bibfield  {author} {\bibinfo {author} {\bibfnamefont {Z.-X.}\ \bibnamefont
			{Liu}}\ and\ \bibinfo {author} {\bibfnamefont {B.}~\bibnamefont {Normand}},\
	}\href {\doibase 10.1103/PhysRevLett.120.187201} {\bibfield  {journal}
		{\bibinfo  {journal} {Phys. Rev. Lett.}\ }\textbf {\bibinfo {volume} {120}},\
		\bibinfo {pages} {187201} (\bibinfo {year} {2018})}\BibitemShut {NoStop}%
	\bibitem [{\citenamefont {Knolle}\ \emph {et~al.}(2014)\citenamefont {Knolle},
		\citenamefont {Chern}, \citenamefont {Kovrizhin}, \citenamefont {Moessner},\
		and\ \citenamefont {Perkins}}]{Raman_Scattering_moessner_2014}%
	\BibitemOpen
	\bibfield  {author} {\bibinfo {author} {\bibfnamefont {J.}~\bibnamefont
			{Knolle}}, \bibinfo {author} {\bibfnamefont {G.-W.}\ \bibnamefont {Chern}},
		\bibinfo {author} {\bibfnamefont {D.~L.}\ \bibnamefont {Kovrizhin}}, \bibinfo
		{author} {\bibfnamefont {R.}~\bibnamefont {Moessner}}, \ and\ \bibinfo
		{author} {\bibfnamefont {N.~B.}\ \bibnamefont {Perkins}},\ }\href {\doibase
		10.1103/PhysRevLett.113.187201} {\bibfield  {journal} {\bibinfo  {journal}
			{Phys. Rev. Lett.}\ }\textbf {\bibinfo {volume} {113}},\ \bibinfo {pages}
		{187201} (\bibinfo {year} {2014})}\BibitemShut {NoStop}%
	\bibitem [{\citenamefont {Hickey}\ and\ \citenamefont
		{Trebst}(2019)}]{hickey2019emergence}%
	\BibitemOpen
	\bibfield  {author} {\bibinfo {author} {\bibfnamefont {C.}~\bibnamefont
			{Hickey}}\ and\ \bibinfo {author} {\bibfnamefont {S.}~\bibnamefont
			{Trebst}},\ }\href@noop {} {\bibfield  {journal} {\bibinfo  {journal} {Nature
				communications}\ }\textbf {\bibinfo {volume} {10}},\ \bibinfo {pages} {530}
		(\bibinfo {year} {2019})}\BibitemShut {NoStop}%
	\bibitem [{\citenamefont {Banerjee}\ \emph {et~al.}(2016)\citenamefont
		{Banerjee}, \citenamefont {Bridges}, \citenamefont {Yan}, \citenamefont
		{Aczel}, \citenamefont {Li}, \citenamefont {Stone}, \citenamefont {Granroth},
		\citenamefont {Lumsden}, \citenamefont {Yiu}, \citenamefont {Knolle} \emph
		{et~al.}}]{banerjee2016}%
	\BibitemOpen
	\bibfield  {author} {\bibinfo {author} {\bibfnamefont {A.}~\bibnamefont
			{Banerjee}}, \bibinfo {author} {\bibfnamefont {C.}~\bibnamefont {Bridges}},
		\bibinfo {author} {\bibfnamefont {J.-Q.}\ \bibnamefont {Yan}}, \bibinfo
		{author} {\bibfnamefont {A.}~\bibnamefont {Aczel}}, \bibinfo {author}
		{\bibfnamefont {L.}~\bibnamefont {Li}}, \bibinfo {author} {\bibfnamefont
			{M.}~\bibnamefont {Stone}}, \bibinfo {author} {\bibfnamefont
			{G.}~\bibnamefont {Granroth}}, \bibinfo {author} {\bibfnamefont
			{M.}~\bibnamefont {Lumsden}}, \bibinfo {author} {\bibfnamefont
			{Y.}~\bibnamefont {Yiu}}, \bibinfo {author} {\bibfnamefont {J.}~\bibnamefont
			{Knolle}},  \emph {et~al.},\ }\href@noop {} {\bibfield  {journal} {\bibinfo
			{journal} {Nature materials}\ }\textbf {\bibinfo {volume} {15}},\ \bibinfo
		{pages} {733} (\bibinfo {year} {2016})}\BibitemShut {NoStop}%
	\bibitem [{\citenamefont {Do}\ \emph {et~al.}(2017)\citenamefont {Do},
		\citenamefont {Park}, \citenamefont {Yoshitake}, \citenamefont {Nasu},
		\citenamefont {Motome}, \citenamefont {Kwon}, \citenamefont {Adroja},
		\citenamefont {Voneshen}, \citenamefont {Kim}, \citenamefont {Jang} \emph
		{et~al.}}]{do2017majorana}%
	\BibitemOpen
	\bibfield  {author} {\bibinfo {author} {\bibfnamefont {S.-H.}\ \bibnamefont
			{Do}}, \bibinfo {author} {\bibfnamefont {S.-Y.}\ \bibnamefont {Park}},
		\bibinfo {author} {\bibfnamefont {J.}~\bibnamefont {Yoshitake}}, \bibinfo
		{author} {\bibfnamefont {J.}~\bibnamefont {Nasu}}, \bibinfo {author}
		{\bibfnamefont {Y.}~\bibnamefont {Motome}}, \bibinfo {author} {\bibfnamefont
			{Y.~S.}\ \bibnamefont {Kwon}}, \bibinfo {author} {\bibfnamefont
			{D.}~\bibnamefont {Adroja}}, \bibinfo {author} {\bibfnamefont
			{D.}~\bibnamefont {Voneshen}}, \bibinfo {author} {\bibfnamefont
			{K.}~\bibnamefont {Kim}}, \bibinfo {author} {\bibfnamefont {T.-H.}\
			\bibnamefont {Jang}},  \emph {et~al.},\ }\href@noop {} {\bibfield  {journal}
		{\bibinfo  {journal} {Nature Physics}\ }\textbf {\bibinfo {volume} {13}},\
		\bibinfo {pages} {1079} (\bibinfo {year} {2017})}\BibitemShut {NoStop}%
	\bibitem [{\citenamefont {Wang}\ \emph {et~al.}(2017)\citenamefont {Wang},
		\citenamefont {Reschke}, \citenamefont {H\"uvonen}, \citenamefont {Do},
		\citenamefont {Choi}, \citenamefont {Gensch}, \citenamefont {Nagel},
		\citenamefont {R\~o\ om},\ and\ \citenamefont {Loidl}}]{WangTHz2017}%
	\BibitemOpen
	\bibfield  {author} {\bibinfo {author} {\bibfnamefont {Z.}~\bibnamefont
			{Wang}}, \bibinfo {author} {\bibfnamefont {S.}~\bibnamefont {Reschke}},
		\bibinfo {author} {\bibfnamefont {D.}~\bibnamefont {H\"uvonen}}, \bibinfo
		{author} {\bibfnamefont {S.-H.}\ \bibnamefont {Do}}, \bibinfo {author}
		{\bibfnamefont {K.-Y.}\ \bibnamefont {Choi}}, \bibinfo {author}
		{\bibfnamefont {M.}~\bibnamefont {Gensch}}, \bibinfo {author} {\bibfnamefont
			{U.}~\bibnamefont {Nagel}}, \bibinfo {author} {\bibfnamefont
			{T.}~\bibnamefont {R\~o\ om}}, \ and\ \bibinfo {author} {\bibfnamefont
			{A.}~\bibnamefont {Loidl}},\ }\href {\doibase 10.1103/PhysRevLett.119.227202}
	{\bibfield  {journal} {\bibinfo  {journal} {Phys. Rev. Lett.}\ }\textbf
		{\bibinfo {volume} {119}},\ \bibinfo {pages} {227202} (\bibinfo {year}
		{2017})}\BibitemShut {NoStop}%
	\bibitem [{\citenamefont {Yu}\ \emph {et~al.}(2018)\citenamefont {Yu},
		\citenamefont {Xu}, \citenamefont {Ran}, \citenamefont {Ni}, \citenamefont
		{Huang}, \citenamefont {Wang}, \citenamefont {Wen},\ and\ \citenamefont
		{Li}}]{yu2018}%
	\BibitemOpen
	\bibfield  {author} {\bibinfo {author} {\bibfnamefont {Y.~J.}\ \bibnamefont
			{Yu}}, \bibinfo {author} {\bibfnamefont {Y.}~\bibnamefont {Xu}}, \bibinfo
		{author} {\bibfnamefont {K.~J.}\ \bibnamefont {Ran}}, \bibinfo {author}
		{\bibfnamefont {J.~M.}\ \bibnamefont {Ni}}, \bibinfo {author} {\bibfnamefont
			{Y.~Y.}\ \bibnamefont {Huang}}, \bibinfo {author} {\bibfnamefont {J.~H.}\
			\bibnamefont {Wang}}, \bibinfo {author} {\bibfnamefont {J.~S.}\ \bibnamefont
			{Wen}}, \ and\ \bibinfo {author} {\bibfnamefont {S.~Y.}\ \bibnamefont {Li}},\
	}\href {\doibase 10.1103/PhysRevLett.120.067202} {\bibfield  {journal}
		{\bibinfo  {journal} {Phys. Rev. Lett.}\ }\textbf {\bibinfo {volume} {120}},\
		\bibinfo {pages} {067202} (\bibinfo {year} {2018})}\BibitemShut {NoStop}%
	\bibitem [{\citenamefont {Singh}\ \emph {et~al.}(2012)\citenamefont {Singh},
		\citenamefont {Manni}, \citenamefont {Reuther}, \citenamefont {Berlijn},
		\citenamefont {Thomale}, \citenamefont {Ku}, \citenamefont {Trebst},\ and\
		\citenamefont {Gegenwart}}]{J_K_model_exp_PhysRevLett.108.127203}%
	\BibitemOpen
	\bibfield  {author} {\bibinfo {author} {\bibfnamefont {Y.}~\bibnamefont
			{Singh}}, \bibinfo {author} {\bibfnamefont {S.}~\bibnamefont {Manni}},
		\bibinfo {author} {\bibfnamefont {J.}~\bibnamefont {Reuther}}, \bibinfo
		{author} {\bibfnamefont {T.}~\bibnamefont {Berlijn}}, \bibinfo {author}
		{\bibfnamefont {R.}~\bibnamefont {Thomale}}, \bibinfo {author} {\bibfnamefont
			{W.}~\bibnamefont {Ku}}, \bibinfo {author} {\bibfnamefont {S.}~\bibnamefont
			{Trebst}}, \ and\ \bibinfo {author} {\bibfnamefont {P.}~\bibnamefont
			{Gegenwart}},\ }\href {\doibase 10.1103/PhysRevLett.108.127203} {\bibfield
		{journal} {\bibinfo  {journal} {Phys. Rev. Lett.}\ }\textbf {\bibinfo
			{volume} {108}},\ \bibinfo {pages} {127203} (\bibinfo {year}
		{2012})}\BibitemShut {NoStop}%
	\bibitem [{\citenamefont {Ye}\ \emph {et~al.}(2012)\citenamefont {Ye},
		\citenamefont {Chi}, \citenamefont {Cao}, \citenamefont {Chakoumakos},
		\citenamefont {Fernandez-Baca}, \citenamefont {Custelcean}, \citenamefont
		{Qi}, \citenamefont {Korneta},\ and\ \citenamefont
		{Cao}}]{zigzag_phase_iridates_PhysRevB.85.180403}%
	\BibitemOpen
	\bibfield  {author} {\bibinfo {author} {\bibfnamefont {F.}~\bibnamefont
			{Ye}}, \bibinfo {author} {\bibfnamefont {S.}~\bibnamefont {Chi}}, \bibinfo
		{author} {\bibfnamefont {H.}~\bibnamefont {Cao}}, \bibinfo {author}
		{\bibfnamefont {B.~C.}\ \bibnamefont {Chakoumakos}}, \bibinfo {author}
		{\bibfnamefont {J.~A.}\ \bibnamefont {Fernandez-Baca}}, \bibinfo {author}
		{\bibfnamefont {R.}~\bibnamefont {Custelcean}}, \bibinfo {author}
		{\bibfnamefont {T.~F.}\ \bibnamefont {Qi}}, \bibinfo {author} {\bibfnamefont
			{O.~B.}\ \bibnamefont {Korneta}}, \ and\ \bibinfo {author} {\bibfnamefont
			{G.}~\bibnamefont {Cao}},\ }\href {\doibase 10.1103/PhysRevB.85.180403}
	{\bibfield  {journal} {\bibinfo  {journal} {Phys. Rev. B}\ }\textbf {\bibinfo
			{volume} {85}},\ \bibinfo {pages} {180403} (\bibinfo {year}
		{2012})}\BibitemShut {NoStop}%
	\bibitem [{\citenamefont {Choi}\ \emph {et~al.}(2012)\citenamefont {Choi},
		\citenamefont {Coldea}, \citenamefont {Kolmogorov}, \citenamefont
		{Lancaster}, \citenamefont {Mazin}, \citenamefont {Blundell}, \citenamefont
		{Radaelli}, \citenamefont {Singh}, \citenamefont {Gegenwart}, \citenamefont
		{Choi}, \citenamefont {Cheong}, \citenamefont {Baker}, \citenamefont
		{Stock},\ and\ \citenamefont
		{Taylor}}]{Spin_waves_Choi_PhysRevLett.108.127204}%
	\BibitemOpen
	\bibfield  {author} {\bibinfo {author} {\bibfnamefont {S.~K.}\ \bibnamefont
			{Choi}}, \bibinfo {author} {\bibfnamefont {R.}~\bibnamefont {Coldea}},
		\bibinfo {author} {\bibfnamefont {A.~N.}\ \bibnamefont {Kolmogorov}},
		\bibinfo {author} {\bibfnamefont {T.}~\bibnamefont {Lancaster}}, \bibinfo
		{author} {\bibfnamefont {I.~I.}\ \bibnamefont {Mazin}}, \bibinfo {author}
		{\bibfnamefont {S.~J.}\ \bibnamefont {Blundell}}, \bibinfo {author}
		{\bibfnamefont {P.~G.}\ \bibnamefont {Radaelli}}, \bibinfo {author}
		{\bibfnamefont {Y.}~\bibnamefont {Singh}}, \bibinfo {author} {\bibfnamefont
			{P.}~\bibnamefont {Gegenwart}}, \bibinfo {author} {\bibfnamefont {K.~R.}\
			\bibnamefont {Choi}}, \bibinfo {author} {\bibfnamefont {S.-W.}\ \bibnamefont
			{Cheong}}, \bibinfo {author} {\bibfnamefont {P.~J.}\ \bibnamefont {Baker}},
		\bibinfo {author} {\bibfnamefont {C.}~\bibnamefont {Stock}}, \ and\ \bibinfo
		{author} {\bibfnamefont {J.}~\bibnamefont {Taylor}},\ }\href {\doibase
		10.1103/PhysRevLett.108.127204} {\bibfield  {journal} {\bibinfo  {journal}
			{Phys. Rev. Lett.}\ }\textbf {\bibinfo {volume} {108}},\ \bibinfo {pages}
		{127204} (\bibinfo {year} {2012})}\BibitemShut {NoStop}%
	\bibitem [{\citenamefont {Singh}\ and\ \citenamefont
		{Gegenwart}(2010)}]{AFM_iridates_PhysRevB.82.064412}%
	\BibitemOpen
	\bibfield  {author} {\bibinfo {author} {\bibfnamefont {Y.}~\bibnamefont
			{Singh}}\ and\ \bibinfo {author} {\bibfnamefont {P.}~\bibnamefont
			{Gegenwart}},\ }\href {\doibase 10.1103/PhysRevB.82.064412} {\bibfield
		{journal} {\bibinfo  {journal} {Phys. Rev. B}\ }\textbf {\bibinfo {volume}
			{82}},\ \bibinfo {pages} {064412} (\bibinfo {year} {2010})}\BibitemShut
	{NoStop}%
	\bibitem [{\citenamefont {Chun}\ \emph {et~al.}(2015)\citenamefont {Chun},
		\citenamefont {Kim}, \citenamefont {Kim}, \citenamefont {Zheng},
		\citenamefont {Stoumpos}, \citenamefont {Malliakas}, \citenamefont
		{Mitchell}, \citenamefont {Mehlawat}, \citenamefont {Singh}, \citenamefont
		{Choi} \emph {et~al.}}]{chun2015direct}%
	\BibitemOpen
	\bibfield  {author} {\bibinfo {author} {\bibfnamefont {S.~H.}\ \bibnamefont
			{Chun}}, \bibinfo {author} {\bibfnamefont {J.-W.}\ \bibnamefont {Kim}},
		\bibinfo {author} {\bibfnamefont {J.}~\bibnamefont {Kim}}, \bibinfo {author}
		{\bibfnamefont {H.}~\bibnamefont {Zheng}}, \bibinfo {author} {\bibfnamefont
			{C.~C.}\ \bibnamefont {Stoumpos}}, \bibinfo {author} {\bibfnamefont
			{C.}~\bibnamefont {Malliakas}}, \bibinfo {author} {\bibfnamefont
			{J.}~\bibnamefont {Mitchell}}, \bibinfo {author} {\bibfnamefont
			{K.}~\bibnamefont {Mehlawat}}, \bibinfo {author} {\bibfnamefont
			{Y.}~\bibnamefont {Singh}}, \bibinfo {author} {\bibfnamefont
			{Y.}~\bibnamefont {Choi}},  \emph {et~al.},\ }\href@noop {} {\bibfield
		{journal} {\bibinfo  {journal} {Nature Physics}\ }\textbf {\bibinfo {volume}
			{11}},\ \bibinfo {pages} {462} (\bibinfo {year} {2015})}\BibitemShut
	{NoStop}%
	\bibitem [{\citenamefont {Comin}\ \emph {et~al.}(2012)\citenamefont {Comin},
		\citenamefont {Levy}, \citenamefont {Ludbrook}, \citenamefont {Zhu},
		\citenamefont {Veenstra}, \citenamefont {Rosen}, \citenamefont {Singh},
		\citenamefont {Gegenwart}, \citenamefont {Stricker}, \citenamefont {Hancock},
		\citenamefont {van~der Marel}, \citenamefont {Elfimov},\ and\ \citenamefont
		{Damascelli}}]{Comin_2012_PhysRevLett.109.266406}%
	\BibitemOpen
	\bibfield  {author} {\bibinfo {author} {\bibfnamefont {R.}~\bibnamefont
			{Comin}}, \bibinfo {author} {\bibfnamefont {G.}~\bibnamefont {Levy}},
		\bibinfo {author} {\bibfnamefont {B.}~\bibnamefont {Ludbrook}}, \bibinfo
		{author} {\bibfnamefont {Z.-H.}\ \bibnamefont {Zhu}}, \bibinfo {author}
		{\bibfnamefont {C.~N.}\ \bibnamefont {Veenstra}}, \bibinfo {author}
		{\bibfnamefont {J.~A.}\ \bibnamefont {Rosen}}, \bibinfo {author}
		{\bibfnamefont {Y.}~\bibnamefont {Singh}}, \bibinfo {author} {\bibfnamefont
			{P.}~\bibnamefont {Gegenwart}}, \bibinfo {author} {\bibfnamefont
			{D.}~\bibnamefont {Stricker}}, \bibinfo {author} {\bibfnamefont {J.~N.}\
			\bibnamefont {Hancock}}, \bibinfo {author} {\bibfnamefont {D.}~\bibnamefont
			{van~der Marel}}, \bibinfo {author} {\bibfnamefont {I.~S.}\ \bibnamefont
			{Elfimov}}, \ and\ \bibinfo {author} {\bibfnamefont {A.}~\bibnamefont
			{Damascelli}},\ }\href {\doibase 10.1103/PhysRevLett.109.266406} {\bibfield
		{journal} {\bibinfo  {journal} {Phys. Rev. Lett.}\ }\textbf {\bibinfo
			{volume} {109}},\ \bibinfo {pages} {266406} (\bibinfo {year}
		{2012})}\BibitemShut {NoStop}%
	\bibitem [{\citenamefont {Gretarsson}\ \emph
		{et~al.}(2013{\natexlab{a}})\citenamefont {Gretarsson}, \citenamefont
		{Clancy}, \citenamefont {Liu}, \citenamefont {Hill}, \citenamefont {Bozin},
		\citenamefont {Singh}, \citenamefont {Manni}, \citenamefont {Gegenwart},
		\citenamefont {Kim}, \citenamefont {Said}, \citenamefont {Casa},
		\citenamefont {Gog}, \citenamefont {Upton}, \citenamefont {Kim},
		\citenamefont {Yu}, \citenamefont {Katukuri}, \citenamefont {Hozoi},
		\citenamefont {van~den Brink},\ and\ \citenamefont
		{Kim}}]{gretarsson2013crystal}%
	\BibitemOpen
	\bibfield  {author} {\bibinfo {author} {\bibfnamefont {H.}~\bibnamefont
			{Gretarsson}}, \bibinfo {author} {\bibfnamefont {J.~P.}\ \bibnamefont
			{Clancy}}, \bibinfo {author} {\bibfnamefont {X.}~\bibnamefont {Liu}},
		\bibinfo {author} {\bibfnamefont {J.~P.}\ \bibnamefont {Hill}}, \bibinfo
		{author} {\bibfnamefont {E.}~\bibnamefont {Bozin}}, \bibinfo {author}
		{\bibfnamefont {Y.}~\bibnamefont {Singh}}, \bibinfo {author} {\bibfnamefont
			{S.}~\bibnamefont {Manni}}, \bibinfo {author} {\bibfnamefont
			{P.}~\bibnamefont {Gegenwart}}, \bibinfo {author} {\bibfnamefont
			{J.}~\bibnamefont {Kim}}, \bibinfo {author} {\bibfnamefont {A.~H.}\
			\bibnamefont {Said}}, \bibinfo {author} {\bibfnamefont {D.}~\bibnamefont
			{Casa}}, \bibinfo {author} {\bibfnamefont {T.}~\bibnamefont {Gog}}, \bibinfo
		{author} {\bibfnamefont {M.~H.}\ \bibnamefont {Upton}}, \bibinfo {author}
		{\bibfnamefont {H.-S.}\ \bibnamefont {Kim}}, \bibinfo {author} {\bibfnamefont
			{J.}~\bibnamefont {Yu}}, \bibinfo {author} {\bibfnamefont {V.~M.}\
			\bibnamefont {Katukuri}}, \bibinfo {author} {\bibfnamefont {L.}~\bibnamefont
			{Hozoi}}, \bibinfo {author} {\bibfnamefont {J.}~\bibnamefont {van~den
				Brink}}, \ and\ \bibinfo {author} {\bibfnamefont {Y.-J.}\ \bibnamefont
			{Kim}},\ }\href {\doibase 10.1103/PhysRevLett.110.076402} {\bibfield
		{journal} {\bibinfo  {journal} {Phys. Rev. Lett.}\ }\textbf {\bibinfo
			{volume} {110}},\ \bibinfo {pages} {076402} (\bibinfo {year}
		{2013}{\natexlab{a}})}\BibitemShut {NoStop}%
	\bibitem [{\citenamefont {Gretarsson}\ \emph
		{et~al.}(2013{\natexlab{b}})\citenamefont {Gretarsson}, \citenamefont
		{Clancy}, \citenamefont {Singh}, \citenamefont {Gegenwart}, \citenamefont
		{Hill}, \citenamefont {Kim}, \citenamefont {Upton}, \citenamefont {Said},
		\citenamefont {Casa}, \citenamefont {Gog},\ and\ \citenamefont
		{Kim}}]{gretarsson2013magnetic}%
	\BibitemOpen
	\bibfield  {author} {\bibinfo {author} {\bibfnamefont {H.}~\bibnamefont
			{Gretarsson}}, \bibinfo {author} {\bibfnamefont {J.~P.}\ \bibnamefont
			{Clancy}}, \bibinfo {author} {\bibfnamefont {Y.}~\bibnamefont {Singh}},
		\bibinfo {author} {\bibfnamefont {P.}~\bibnamefont {Gegenwart}}, \bibinfo
		{author} {\bibfnamefont {J.~P.}\ \bibnamefont {Hill}}, \bibinfo {author}
		{\bibfnamefont {J.}~\bibnamefont {Kim}}, \bibinfo {author} {\bibfnamefont
			{M.~H.}\ \bibnamefont {Upton}}, \bibinfo {author} {\bibfnamefont {A.~H.}\
			\bibnamefont {Said}}, \bibinfo {author} {\bibfnamefont {D.}~\bibnamefont
			{Casa}}, \bibinfo {author} {\bibfnamefont {T.}~\bibnamefont {Gog}}, \ and\
		\bibinfo {author} {\bibfnamefont {Y.-J.}\ \bibnamefont {Kim}},\ }\href
	{\doibase 10.1103/PhysRevB.87.220407} {\bibfield  {journal} {\bibinfo
			{journal} {Phys. Rev. B}\ }\textbf {\bibinfo {volume} {87}},\ \bibinfo
		{pages} {220407} (\bibinfo {year} {2013}{\natexlab{b}})}\BibitemShut
	{NoStop}%
	\bibitem [{\citenamefont {Liu}\ \emph {et~al.}(2011)\citenamefont {Liu},
		\citenamefont {Berlijn}, \citenamefont {Yin}, \citenamefont {Ku},
		\citenamefont {Tsvelik}, \citenamefont {Kim}, \citenamefont {Gretarsson},
		\citenamefont {Singh}, \citenamefont {Gegenwart},\ and\ \citenamefont
		{Hill}}]{liu2011long}%
	\BibitemOpen
	\bibfield  {author} {\bibinfo {author} {\bibfnamefont {X.}~\bibnamefont
			{Liu}}, \bibinfo {author} {\bibfnamefont {T.}~\bibnamefont {Berlijn}},
		\bibinfo {author} {\bibfnamefont {W.-G.}\ \bibnamefont {Yin}}, \bibinfo
		{author} {\bibfnamefont {W.}~\bibnamefont {Ku}}, \bibinfo {author}
		{\bibfnamefont {A.}~\bibnamefont {Tsvelik}}, \bibinfo {author} {\bibfnamefont
			{Y.-J.}\ \bibnamefont {Kim}}, \bibinfo {author} {\bibfnamefont
			{H.}~\bibnamefont {Gretarsson}}, \bibinfo {author} {\bibfnamefont
			{Y.}~\bibnamefont {Singh}}, \bibinfo {author} {\bibfnamefont
			{P.}~\bibnamefont {Gegenwart}}, \ and\ \bibinfo {author} {\bibfnamefont
			{J.~P.}\ \bibnamefont {Hill}},\ }\href {\doibase 10.1103/PhysRevB.83.220403}
	{\bibfield  {journal} {\bibinfo  {journal} {Phys. Rev. B}\ }\textbf {\bibinfo
			{volume} {83}},\ \bibinfo {pages} {220403} (\bibinfo {year}
		{2011})}\BibitemShut {NoStop}%
	\bibitem [{\citenamefont {Mehlawat}\ \emph {et~al.}(2017)\citenamefont
		{Mehlawat}, \citenamefont {Thamizhavel},\ and\ \citenamefont
		{Singh}}]{mehlawat2017heat}%
	\BibitemOpen
	\bibfield  {author} {\bibinfo {author} {\bibfnamefont {K.}~\bibnamefont
			{Mehlawat}}, \bibinfo {author} {\bibfnamefont {A.}~\bibnamefont
			{Thamizhavel}}, \ and\ \bibinfo {author} {\bibfnamefont {Y.}~\bibnamefont
			{Singh}},\ }\href {\doibase 10.1103/PhysRevB.95.144406} {\bibfield  {journal}
		{\bibinfo  {journal} {Phys. Rev. B}\ }\textbf {\bibinfo {volume} {95}},\
		\bibinfo {pages} {144406} (\bibinfo {year} {2017})}\BibitemShut {NoStop}%
	\bibitem [{\citenamefont {Banerjee}\ \emph {et~al.}(2018)\citenamefont
		{Banerjee}, \citenamefont {Lampen-Kelley}, \citenamefont {Knolle},
		\citenamefont {Balz}, \citenamefont {Aczel}, \citenamefont {Winn},
		\citenamefont {Liu}, \citenamefont {Pajerowski}, \citenamefont {Yan},
		\citenamefont {Bridges} \emph {et~al.}}]{banerjee2018excitations}%
	\BibitemOpen
	\bibfield  {author} {\bibinfo {author} {\bibfnamefont {A.}~\bibnamefont
			{Banerjee}}, \bibinfo {author} {\bibfnamefont {P.}~\bibnamefont
			{Lampen-Kelley}}, \bibinfo {author} {\bibfnamefont {J.}~\bibnamefont
			{Knolle}}, \bibinfo {author} {\bibfnamefont {C.}~\bibnamefont {Balz}},
		\bibinfo {author} {\bibfnamefont {A.~A.}\ \bibnamefont {Aczel}}, \bibinfo
		{author} {\bibfnamefont {B.}~\bibnamefont {Winn}}, \bibinfo {author}
		{\bibfnamefont {Y.}~\bibnamefont {Liu}}, \bibinfo {author} {\bibfnamefont
			{D.}~\bibnamefont {Pajerowski}}, \bibinfo {author} {\bibfnamefont
			{J.}~\bibnamefont {Yan}}, \bibinfo {author} {\bibfnamefont {C.~A.}\
			\bibnamefont {Bridges}},  \emph {et~al.},\ }\href@noop {} {\bibfield
		{journal} {\bibinfo  {journal} {npj Quantum Materials}\ }\textbf {\bibinfo
			{volume} {3}},\ \bibinfo {pages} {8} (\bibinfo {year} {2018})}\BibitemShut
	{NoStop}%
	\bibitem [{\citenamefont {Wu}\ \emph {et~al.}(2018)\citenamefont {Wu},
		\citenamefont {Little}, \citenamefont {Aldape}, \citenamefont {Rees},
		\citenamefont {Thewalt}, \citenamefont {Lampen-Kelley}, \citenamefont
		{Banerjee}, \citenamefont {Bridges}, \citenamefont {Yan}, \citenamefont
		{Boone}, \citenamefont {Patankar}, \citenamefont {Goldhaber-Gordon},
		\citenamefont {Mandrus}, \citenamefont {Nagler}, \citenamefont {Altman},\
		and\ \citenamefont {Orenstein}}]{magnons_THZ}%
	\BibitemOpen
	\bibfield  {author} {\bibinfo {author} {\bibfnamefont {L.}~\bibnamefont
			{Wu}}, \bibinfo {author} {\bibfnamefont {A.}~\bibnamefont {Little}}, \bibinfo
		{author} {\bibfnamefont {E.~E.}\ \bibnamefont {Aldape}}, \bibinfo {author}
		{\bibfnamefont {D.}~\bibnamefont {Rees}}, \bibinfo {author} {\bibfnamefont
			{E.}~\bibnamefont {Thewalt}}, \bibinfo {author} {\bibfnamefont
			{P.}~\bibnamefont {Lampen-Kelley}}, \bibinfo {author} {\bibfnamefont
			{A.}~\bibnamefont {Banerjee}}, \bibinfo {author} {\bibfnamefont {C.~A.}\
			\bibnamefont {Bridges}}, \bibinfo {author} {\bibfnamefont {J.-Q.}\
			\bibnamefont {Yan}}, \bibinfo {author} {\bibfnamefont {D.}~\bibnamefont
			{Boone}}, \bibinfo {author} {\bibfnamefont {S.}~\bibnamefont {Patankar}},
		\bibinfo {author} {\bibfnamefont {D.}~\bibnamefont {Goldhaber-Gordon}},
		\bibinfo {author} {\bibfnamefont {D.}~\bibnamefont {Mandrus}}, \bibinfo
		{author} {\bibfnamefont {S.~E.}\ \bibnamefont {Nagler}}, \bibinfo {author}
		{\bibfnamefont {E.}~\bibnamefont {Altman}}, \ and\ \bibinfo {author}
		{\bibfnamefont {J.}~\bibnamefont {Orenstein}},\ }\href {\doibase
		10.1103/PhysRevB.98.094425} {\bibfield  {journal} {\bibinfo  {journal} {Phys.
				Rev. B}\ }\textbf {\bibinfo {volume} {98}},\ \bibinfo {pages} {094425}
		(\bibinfo {year} {2018})}\BibitemShut {NoStop}%
	\bibitem [{\citenamefont {Kasahara}\ \emph {et~al.}(2018)\citenamefont
		{Kasahara}, \citenamefont {Ohnishi}, \citenamefont {Mizukami}, \citenamefont
		{Tanaka}, \citenamefont {Ma}, \citenamefont {Sugii}, \citenamefont {Kurita},
		\citenamefont {Tanaka}, \citenamefont {Nasu}, \citenamefont {Motome} \emph
		{et~al.}}]{matsuda2018majorana}%
	\BibitemOpen
	\bibfield  {author} {\bibinfo {author} {\bibfnamefont {Y.}~\bibnamefont
			{Kasahara}}, \bibinfo {author} {\bibfnamefont {T.}~\bibnamefont {Ohnishi}},
		\bibinfo {author} {\bibfnamefont {Y.}~\bibnamefont {Mizukami}}, \bibinfo
		{author} {\bibfnamefont {O.}~\bibnamefont {Tanaka}}, \bibinfo {author}
		{\bibfnamefont {S.}~\bibnamefont {Ma}}, \bibinfo {author} {\bibfnamefont
			{K.}~\bibnamefont {Sugii}}, \bibinfo {author} {\bibfnamefont
			{N.}~\bibnamefont {Kurita}}, \bibinfo {author} {\bibfnamefont
			{H.}~\bibnamefont {Tanaka}}, \bibinfo {author} {\bibfnamefont
			{J.}~\bibnamefont {Nasu}}, \bibinfo {author} {\bibfnamefont {Y.}~\bibnamefont
			{Motome}},  \emph {et~al.},\ }\href@noop {} {\bibfield  {journal} {\bibinfo
			{journal} {Nature}\ }\textbf {\bibinfo {volume} {559}},\ \bibinfo {pages}
		{227} (\bibinfo {year} {2018})}\BibitemShut {NoStop}%
	\bibitem [{\citenamefont {Ran}\ \emph {et~al.}(2017)\citenamefont {Ran},
		\citenamefont {Wang}, \citenamefont {Wang}, \citenamefont {Dong},
		\citenamefont {Ren}, \citenamefont {Bao}, \citenamefont {Li}, \citenamefont
		{Ma}, \citenamefont {Gan}, \citenamefont {Zhang}, \citenamefont {Park},
		\citenamefont {Deng}, \citenamefont {Danilkin}, \citenamefont {Yu},
		\citenamefont {Li},\ and\ \citenamefont {Wen}}]{inelastic_neutron}%
	\BibitemOpen
	\bibfield  {author} {\bibinfo {author} {\bibfnamefont {K.}~\bibnamefont
			{Ran}}, \bibinfo {author} {\bibfnamefont {J.}~\bibnamefont {Wang}}, \bibinfo
		{author} {\bibfnamefont {W.}~\bibnamefont {Wang}}, \bibinfo {author}
		{\bibfnamefont {Z.-Y.}\ \bibnamefont {Dong}}, \bibinfo {author}
		{\bibfnamefont {X.}~\bibnamefont {Ren}}, \bibinfo {author} {\bibfnamefont
			{S.}~\bibnamefont {Bao}}, \bibinfo {author} {\bibfnamefont {S.}~\bibnamefont
			{Li}}, \bibinfo {author} {\bibfnamefont {Z.}~\bibnamefont {Ma}}, \bibinfo
		{author} {\bibfnamefont {Y.}~\bibnamefont {Gan}}, \bibinfo {author}
		{\bibfnamefont {Y.}~\bibnamefont {Zhang}}, \bibinfo {author} {\bibfnamefont
			{J.~T.}\ \bibnamefont {Park}}, \bibinfo {author} {\bibfnamefont
			{G.}~\bibnamefont {Deng}}, \bibinfo {author} {\bibfnamefont {S.}~\bibnamefont
			{Danilkin}}, \bibinfo {author} {\bibfnamefont {S.-L.}\ \bibnamefont {Yu}},
		\bibinfo {author} {\bibfnamefont {J.-X.}\ \bibnamefont {Li}}, \ and\ \bibinfo
		{author} {\bibfnamefont {J.}~\bibnamefont {Wen}},\ }\href {\doibase
		10.1103/PhysRevLett.118.107203} {\bibfield  {journal} {\bibinfo  {journal}
			{Phys. Rev. Lett.}\ }\textbf {\bibinfo {volume} {118}},\ \bibinfo {pages}
		{107203} (\bibinfo {year} {2017})}\BibitemShut {NoStop}%
	\bibitem [{\citenamefont {Sandilands}\ \emph {et~al.}(2016)\citenamefont
		{Sandilands}, \citenamefont {Tian}, \citenamefont {Reijnders}, \citenamefont
		{Kim}, \citenamefont {Plumb}, \citenamefont {Kim}, \citenamefont {Kee},\ and\
		\citenamefont {Burch}}]{Sandilands_kitaev_candidates}%
	\BibitemOpen
	\bibfield  {author} {\bibinfo {author} {\bibfnamefont {L.~J.}\ \bibnamefont
			{Sandilands}}, \bibinfo {author} {\bibfnamefont {Y.}~\bibnamefont {Tian}},
		\bibinfo {author} {\bibfnamefont {A.~A.}\ \bibnamefont {Reijnders}}, \bibinfo
		{author} {\bibfnamefont {H.-S.}\ \bibnamefont {Kim}}, \bibinfo {author}
		{\bibfnamefont {K.~W.}\ \bibnamefont {Plumb}}, \bibinfo {author}
		{\bibfnamefont {Y.-J.}\ \bibnamefont {Kim}}, \bibinfo {author} {\bibfnamefont
			{H.-Y.}\ \bibnamefont {Kee}}, \ and\ \bibinfo {author} {\bibfnamefont
			{K.~S.}\ \bibnamefont {Burch}},\ }\href {\doibase 10.1103/PhysRevB.93.075144}
	{\bibfield  {journal} {\bibinfo  {journal} {Phys. Rev. B}\ }\textbf {\bibinfo
			{volume} {93}},\ \bibinfo {pages} {075144} (\bibinfo {year}
		{2016})}\BibitemShut {NoStop}%
	\bibitem [{\citenamefont {Sandilands}\ \emph {et~al.}(2015)\citenamefont
		{Sandilands}, \citenamefont {Tian}, \citenamefont {Plumb}, \citenamefont
		{Kim},\ and\ \citenamefont {Burch}}]{Sandilands_luke_kitaev_candidates}%
	\BibitemOpen
	\bibfield  {author} {\bibinfo {author} {\bibfnamefont {L.~J.}\ \bibnamefont
			{Sandilands}}, \bibinfo {author} {\bibfnamefont {Y.}~\bibnamefont {Tian}},
		\bibinfo {author} {\bibfnamefont {K.~W.}\ \bibnamefont {Plumb}}, \bibinfo
		{author} {\bibfnamefont {Y.-J.}\ \bibnamefont {Kim}}, \ and\ \bibinfo
		{author} {\bibfnamefont {K.~S.}\ \bibnamefont {Burch}},\ }\href {\doibase
		10.1103/PhysRevLett.114.147201} {\bibfield  {journal} {\bibinfo  {journal}
			{Phys. Rev. Lett.}\ }\textbf {\bibinfo {volume} {114}},\ \bibinfo {pages}
		{147201} (\bibinfo {year} {2015})}\BibitemShut {NoStop}%
	\bibitem [{\citenamefont {Plumb}\ \emph {et~al.}(2014)\citenamefont {Plumb},
		\citenamefont {Clancy}, \citenamefont {Sandilands}, \citenamefont {Shankar},
		\citenamefont {Hu}, \citenamefont {Burch}, \citenamefont {Kee},\ and\
		\citenamefont {Kim}}]{Plumb_iridates}%
	\BibitemOpen
	\bibfield  {author} {\bibinfo {author} {\bibfnamefont {K.~W.}\ \bibnamefont
			{Plumb}}, \bibinfo {author} {\bibfnamefont {J.~P.}\ \bibnamefont {Clancy}},
		\bibinfo {author} {\bibfnamefont {L.~J.}\ \bibnamefont {Sandilands}},
		\bibinfo {author} {\bibfnamefont {V.~V.}\ \bibnamefont {Shankar}}, \bibinfo
		{author} {\bibfnamefont {Y.~F.}\ \bibnamefont {Hu}}, \bibinfo {author}
		{\bibfnamefont {K.~S.}\ \bibnamefont {Burch}}, \bibinfo {author}
		{\bibfnamefont {H.-Y.}\ \bibnamefont {Kee}}, \ and\ \bibinfo {author}
		{\bibfnamefont {Y.-J.}\ \bibnamefont {Kim}},\ }\href {\doibase
		10.1103/PhysRevB.90.041112} {\bibfield  {journal} {\bibinfo  {journal} {Phys.
				Rev. B}\ }\textbf {\bibinfo {volume} {90}},\ \bibinfo {pages} {041112}
		(\bibinfo {year} {2014})}\BibitemShut {NoStop}%
	\bibitem [{\citenamefont {Jackeli}\ and\ \citenamefont
		{Khaliullin}(2009)}]{Jackeli_Mott}%
	\BibitemOpen
	\bibfield  {author} {\bibinfo {author} {\bibfnamefont {G.}~\bibnamefont
			{Jackeli}}\ and\ \bibinfo {author} {\bibfnamefont {G.}~\bibnamefont
			{Khaliullin}},\ }\href {\doibase 10.1103/PhysRevLett.102.017205} {\bibfield
		{journal} {\bibinfo  {journal} {Phys. Rev. Lett.}\ }\textbf {\bibinfo
			{volume} {102}},\ \bibinfo {pages} {017205} (\bibinfo {year}
		{2009})}\BibitemShut {NoStop}%
	\bibitem [{\citenamefont {Leahy}\ \emph {et~al.}(2017)\citenamefont {Leahy},
		\citenamefont {Pocs}, \citenamefont {Siegfried}, \citenamefont {Graf},
		\citenamefont {Do}, \citenamefont {Choi}, \citenamefont {Normand},\ and\
		\citenamefont {Lee}}]{Leahy2017_thermalconductivity_torque}%
	\BibitemOpen
	\bibfield  {author} {\bibinfo {author} {\bibfnamefont {I.~A.}\ \bibnamefont
			{Leahy}}, \bibinfo {author} {\bibfnamefont {C.~A.}\ \bibnamefont {Pocs}},
		\bibinfo {author} {\bibfnamefont {P.~E.}\ \bibnamefont {Siegfried}}, \bibinfo
		{author} {\bibfnamefont {D.}~\bibnamefont {Graf}}, \bibinfo {author}
		{\bibfnamefont {S.-H.}\ \bibnamefont {Do}}, \bibinfo {author} {\bibfnamefont
			{K.-Y.}\ \bibnamefont {Choi}}, \bibinfo {author} {\bibfnamefont
			{B.}~\bibnamefont {Normand}}, \ and\ \bibinfo {author} {\bibfnamefont
			{M.}~\bibnamefont {Lee}},\ }\href {\doibase 10.1103/PhysRevLett.118.187203}
	{\bibfield  {journal} {\bibinfo  {journal} {Phys. Rev. Lett.}\ }\textbf
		{\bibinfo {volume} {118}},\ \bibinfo {pages} {187203} (\bibinfo {year}
		{2017})}\BibitemShut {NoStop}%
	\bibitem [{\citenamefont {Das}\ \emph {et~al.}(2019)\citenamefont {Das},
		\citenamefont {Kundu}, \citenamefont {Zhu}, \citenamefont {Mun},
		\citenamefont {McDonald}, \citenamefont {Li}, \citenamefont {Balicas},
		\citenamefont {McCollam}, \citenamefont {Cao}, \citenamefont {Rau},
		\citenamefont {Kee}, \citenamefont {Tripathi},\ and\ \citenamefont
		{Sebastian}}]{torque_responce}%
	\BibitemOpen
	\bibfield  {author} {\bibinfo {author} {\bibfnamefont {S.~D.}\ \bibnamefont
			{Das}}, \bibinfo {author} {\bibfnamefont {S.}~\bibnamefont {Kundu}}, \bibinfo
		{author} {\bibfnamefont {Z.}~\bibnamefont {Zhu}}, \bibinfo {author}
		{\bibfnamefont {E.}~\bibnamefont {Mun}}, \bibinfo {author} {\bibfnamefont
			{R.~D.}\ \bibnamefont {McDonald}}, \bibinfo {author} {\bibfnamefont
			{G.}~\bibnamefont {Li}}, \bibinfo {author} {\bibfnamefont {L.}~\bibnamefont
			{Balicas}}, \bibinfo {author} {\bibfnamefont {A.}~\bibnamefont {McCollam}},
		\bibinfo {author} {\bibfnamefont {G.}~\bibnamefont {Cao}}, \bibinfo {author}
		{\bibfnamefont {J.~G.}\ \bibnamefont {Rau}}, \bibinfo {author} {\bibfnamefont
			{H.-Y.}\ \bibnamefont {Kee}}, \bibinfo {author} {\bibfnamefont
			{V.}~\bibnamefont {Tripathi}}, \ and\ \bibinfo {author} {\bibfnamefont
			{S.~E.}\ \bibnamefont {Sebastian}},\ }\href {\doibase
		10.1103/PhysRevB.99.081101} {\bibfield  {journal} {\bibinfo  {journal} {Phys.
				Rev. B}\ }\textbf {\bibinfo {volume} {99}},\ \bibinfo {pages} {081101}
		(\bibinfo {year} {2019})}\BibitemShut {NoStop}%
	\bibitem [{\citenamefont {Baek}\ \emph {et~al.}(2017)\citenamefont {Baek},
		\citenamefont {Do}, \citenamefont {Choi}, \citenamefont {Kwon}, \citenamefont
		{Wolter}, \citenamefont {Nishimoto}, \citenamefont {van~den Brink},\ and\
		\citenamefont {B\"uchner}}]{baek2017evidence}%
	\BibitemOpen
	\bibfield  {author} {\bibinfo {author} {\bibfnamefont {S.-H.}\ \bibnamefont
			{Baek}}, \bibinfo {author} {\bibfnamefont {S.-H.}\ \bibnamefont {Do}},
		\bibinfo {author} {\bibfnamefont {K.-Y.}\ \bibnamefont {Choi}}, \bibinfo
		{author} {\bibfnamefont {Y.~S.}\ \bibnamefont {Kwon}}, \bibinfo {author}
		{\bibfnamefont {A.~U.~B.}\ \bibnamefont {Wolter}}, \bibinfo {author}
		{\bibfnamefont {S.}~\bibnamefont {Nishimoto}}, \bibinfo {author}
		{\bibfnamefont {J.}~\bibnamefont {van~den Brink}}, \ and\ \bibinfo {author}
		{\bibfnamefont {B.}~\bibnamefont {B\"uchner}},\ }\href {\doibase
		10.1103/PhysRevLett.119.037201} {\bibfield  {journal} {\bibinfo  {journal}
			{Phys. Rev. Lett.}\ }\textbf {\bibinfo {volume} {119}},\ \bibinfo {pages}
		{037201} (\bibinfo {year} {2017})}\BibitemShut {NoStop}%
	\bibitem [{\citenamefont {Polizzi}(2009)}]{FEAST}%
	\BibitemOpen
	\bibfield  {author} {\bibinfo {author} {\bibfnamefont {E.}~\bibnamefont
			{Polizzi}},\ }\href {\doibase 10.1103/PhysRevB.79.115112} {\bibfield
		{journal} {\bibinfo  {journal} {Phys. Rev. B}\ }\textbf {\bibinfo {volume}
			{79}},\ \bibinfo {pages} {115112} (\bibinfo {year} {2009})}\BibitemShut
	{NoStop}%
	\bibitem [{\citenamefont {Stewart}(2002)}]{krylov_schur}%
	\BibitemOpen
	\bibfield  {author} {\bibinfo {author} {\bibfnamefont {G.~W.}\ \bibnamefont
			{Stewart}},\ }\href@noop {} {\bibfield  {journal} {\bibinfo  {journal} {SIAM
				Journal on Matrix Analysis and Applications}\ }\textbf {\bibinfo {volume}
			{23}},\ \bibinfo {pages} {601} (\bibinfo {year} {2002})}\BibitemShut
	{NoStop}%
	\bibitem [{\citenamefont {Smith}\ \emph
		{et~al.}(2017{\natexlab{a}})\citenamefont {Smith}, \citenamefont {Knolle},
		\citenamefont {Kovrizhin},\ and\ \citenamefont
		{Moessner}}]{Moessner_disorder_free_localization}%
	\BibitemOpen
	\bibfield  {author} {\bibinfo {author} {\bibfnamefont {A.}~\bibnamefont
			{Smith}}, \bibinfo {author} {\bibfnamefont {J.}~\bibnamefont {Knolle}},
		\bibinfo {author} {\bibfnamefont {D.~L.}\ \bibnamefont {Kovrizhin}}, \ and\
		\bibinfo {author} {\bibfnamefont {R.}~\bibnamefont {Moessner}},\ }\href
	{\doibase 10.1103/PhysRevLett.118.266601} {\bibfield  {journal} {\bibinfo
			{journal} {Phys. Rev. Lett.}\ }\textbf {\bibinfo {volume} {118}},\ \bibinfo
		{pages} {266601} (\bibinfo {year} {2017}{\natexlab{a}})}\BibitemShut
	{NoStop}%
	\bibitem [{\citenamefont {Smith}\ \emph
		{et~al.}(2017{\natexlab{b}})\citenamefont {Smith}, \citenamefont {Knolle},
		\citenamefont {Moessner},\ and\ \citenamefont
		{Kovrizhin}}]{Moessner_Absence_ergodicity}%
	\BibitemOpen
	\bibfield  {author} {\bibinfo {author} {\bibfnamefont {A.}~\bibnamefont
			{Smith}}, \bibinfo {author} {\bibfnamefont {J.}~\bibnamefont {Knolle}},
		\bibinfo {author} {\bibfnamefont {R.}~\bibnamefont {Moessner}}, \ and\
		\bibinfo {author} {\bibfnamefont {D.~L.}\ \bibnamefont {Kovrizhin}},\ }\href
	{\doibase 10.1103/PhysRevLett.119.176601} {\bibfield  {journal} {\bibinfo
			{journal} {Phys. Rev. Lett.}\ }\textbf {\bibinfo {volume} {119}},\ \bibinfo
		{pages} {176601} (\bibinfo {year} {2017}{\natexlab{b}})}\BibitemShut
	{NoStop}%
	\bibitem [{\citenamefont {Brenes}\ \emph {et~al.}(2018)\citenamefont {Brenes},
		\citenamefont {Dalmonte}, \citenamefont {Heyl},\ and\ \citenamefont
		{Scardicchio}}]{Brenes_gauge_invariance_mbl}%
	\BibitemOpen
	\bibfield  {author} {\bibinfo {author} {\bibfnamefont {M.}~\bibnamefont
			{Brenes}}, \bibinfo {author} {\bibfnamefont {M.}~\bibnamefont {Dalmonte}},
		\bibinfo {author} {\bibfnamefont {M.}~\bibnamefont {Heyl}}, \ and\ \bibinfo
		{author} {\bibfnamefont {A.}~\bibnamefont {Scardicchio}},\ }\href {\doibase
		10.1103/PhysRevLett.120.030601} {\bibfield  {journal} {\bibinfo  {journal}
			{Phys. Rev. Lett.}\ }\textbf {\bibinfo {volume} {120}},\ \bibinfo {pages}
		{030601} (\bibinfo {year} {2018})}\BibitemShut {NoStop}%
	\bibitem [{\citenamefont {Diamantini}\ \emph {et~al.}(2018)\citenamefont
		{Diamantini}, \citenamefont {Trugenberger},\ and\ \citenamefont
		{Vinokur}}]{diamantini2018confinement}%
	\BibitemOpen
	\bibfield  {author} {\bibinfo {author} {\bibfnamefont {M.~C.}\ \bibnamefont
			{Diamantini}}, \bibinfo {author} {\bibfnamefont {C.~A.}\ \bibnamefont
			{Trugenberger}}, \ and\ \bibinfo {author} {\bibfnamefont {V.~M.}\
			\bibnamefont {Vinokur}},\ }\href@noop {} {\bibfield  {journal} {\bibinfo
			{journal} {Communications Physics}\ }\textbf {\bibinfo {volume} {1}},\
		\bibinfo {pages} {77} (\bibinfo {year} {2018})}\BibitemShut {NoStop}%
	\bibitem [{\citenamefont {Altshuler}\ \emph {et~al.}(1997)\citenamefont
		{Altshuler}, \citenamefont {Gefen}, \citenamefont {Kamenev},\ and\
		\citenamefont {Levitov}}]{altshuler1997}%
	\BibitemOpen
	\bibfield  {author} {\bibinfo {author} {\bibfnamefont {B.~L.}\ \bibnamefont
			{Altshuler}}, \bibinfo {author} {\bibfnamefont {Y.}~\bibnamefont {Gefen}},
		\bibinfo {author} {\bibfnamefont {A.}~\bibnamefont {Kamenev}}, \ and\
		\bibinfo {author} {\bibfnamefont {L.~S.}\ \bibnamefont {Levitov}},\ }\href
	{\doibase 10.1103/PhysRevLett.78.2803} {\bibfield  {journal} {\bibinfo
			{journal} {Phys. Rev. Lett.}\ }\textbf {\bibinfo {volume} {78}},\ \bibinfo
		{pages} {2803} (\bibinfo {year} {1997})}\BibitemShut {NoStop}%
	\bibitem [{\citenamefont {Mirlin}\ and\ \citenamefont
		{Fyodorov}(1997)}]{mirlin1997}%
	\BibitemOpen
	\bibfield  {author} {\bibinfo {author} {\bibfnamefont {A.~D.}\ \bibnamefont
			{Mirlin}}\ and\ \bibinfo {author} {\bibfnamefont {Y.~V.}\ \bibnamefont
			{Fyodorov}},\ }\href {\doibase 10.1103/PhysRevB.56.13393} {\bibfield
		{journal} {\bibinfo  {journal} {Phys. Rev. B}\ }\textbf {\bibinfo {volume}
			{56}},\ \bibinfo {pages} {13393} (\bibinfo {year} {1997})}\BibitemShut
	{NoStop}%
	\bibitem [{\citenamefont {Rousochatzakis}\ \emph {et~al.}(2019)\citenamefont
		{Rousochatzakis}, \citenamefont {Kourtis}, \citenamefont {Knolle},
		\citenamefont {Moessner},\ and\ \citenamefont {Perkins}}]{Rousochatzakis}%
	\BibitemOpen
	\bibfield  {author} {\bibinfo {author} {\bibfnamefont {I.}~\bibnamefont
			{Rousochatzakis}}, \bibinfo {author} {\bibfnamefont {S.}~\bibnamefont
			{Kourtis}}, \bibinfo {author} {\bibfnamefont {J.}~\bibnamefont {Knolle}},
		\bibinfo {author} {\bibfnamefont {R.}~\bibnamefont {Moessner}}, \ and\
		\bibinfo {author} {\bibfnamefont {N.~B.}\ \bibnamefont {Perkins}},\ }\href
	{\doibase 10.1103/PhysRevB.100.045117} {\bibfield  {journal} {\bibinfo
			{journal} {Phys. Rev. B}\ }\textbf {\bibinfo {volume} {100}},\ \bibinfo
		{pages} {045117} (\bibinfo {year} {2019})}\BibitemShut {NoStop}%
	\bibitem [{\citenamefont {Di~Napoli}\ \emph {et~al.}(2016)\citenamefont
		{Di~Napoli}, \citenamefont {Polizzi},\ and\ \citenamefont
		{Saad}}]{di2016efficient}%
	\BibitemOpen
	\bibfield  {author} {\bibinfo {author} {\bibfnamefont {E.}~\bibnamefont
			{Di~Napoli}}, \bibinfo {author} {\bibfnamefont {E.}~\bibnamefont {Polizzi}},
		\ and\ \bibinfo {author} {\bibfnamefont {Y.}~\bibnamefont {Saad}},\
	}\href@noop {} {\bibfield  {journal} {\bibinfo  {journal} {Numerical Linear
				Algebra with Applications}\ }\textbf {\bibinfo {volume} {23}},\ \bibinfo
		{pages} {674} (\bibinfo {year} {2016})}\BibitemShut {NoStop}%
	\bibitem [{\citenamefont {Sandvik}(2010)}]{sandvik2010computational}%
	\BibitemOpen
	\bibfield  {author} {\bibinfo {author} {\bibfnamefont {A.~W.}\ \bibnamefont
			{Sandvik}},\ }in\ \href@noop {} {\emph {\bibinfo {booktitle} {AIP Conference
				Proceedings}}},\ Vol.\ \bibinfo {volume} {1297}\ (\bibinfo {organization}
	{American Institute of Physics},\ \bibinfo {year} {2010})\ pp.\ \bibinfo
	{pages} {135--338}\BibitemShut {NoStop}%
	\bibitem [{\citenamefont {Samajdar}\ \emph {et~al.}(2019)\citenamefont
		{Samajdar}, \citenamefont {Scheurer}, \citenamefont {Chatterjee},
		\citenamefont {Guo}, \citenamefont {Xu},\ and\ \citenamefont
		{Sachdev}}]{Sachdevcuprates}%
	\BibitemOpen
	\bibfield  {author} {\bibinfo {author} {\bibfnamefont {R.}~\bibnamefont
			{Samajdar}}, \bibinfo {author} {\bibfnamefont {M.~S.}\ \bibnamefont
			{Scheurer}}, \bibinfo {author} {\bibfnamefont {S.}~\bibnamefont
			{Chatterjee}}, \bibinfo {author} {\bibfnamefont {H.}~\bibnamefont {Guo}},
		\bibinfo {author} {\bibfnamefont {C.}~\bibnamefont {Xu}}, \ and\ \bibinfo
		{author} {\bibfnamefont {S.}~\bibnamefont {Sachdev}},\ }\href@noop {}
	{\bibfield  {journal} {\bibinfo  {journal} {Nature Physics}\ }\textbf
		{\bibinfo {volume} {15}},\ \bibinfo {pages} {1290} (\bibinfo {year}
		{2019})}\BibitemShut {NoStop}%
\end{thebibliography}
\end{document}